\documentclass[11pt]{article}
\usepackage[utf8]{inputenc}
\usepackage[a4paper, total={6in, 8.2in}]{geometry}
\usepackage{amsmath}
\usepackage{mathtools}
\usepackage{amsthm}
\usepackage{amssymb}
\usepackage[english]{ babel}
\usepackage{dirtytalk}
\usepackage{graphicx}
\usepackage{authblk}
\usepackage[hyphens]{url}
\usepackage{hyperref}
\usepackage{caption}
\usepackage{subcaption}
\usepackage{cases}
\usepackage{xcolor} 
\usepackage{bm}
\usepackage{tikz}
\usetikzlibrary{shapes,arrows.meta,positioning}

\usepackage{tabularx}
\usepackage{blkarray}
\usepackage{enumerate}
\usepackage{comment}
\usepackage{algorithm}
\usepackage{algorithmic}
\usepackage{float}
\usepackage{booktabs}
\usepackage{ragged2e}

\usepackage{appendix}
\usepackage{multirow}

\usepackage[bottom]{footmisc}
\title{A Low-Cost IoT Device for Environmental Monitoring and Embedded Solar Forecasting with On-Device Incremental Learning}

\author[1]{Erick Michel Lara Pinal}
\author[2]{Abhinav Das}
\author[3]{Stephan Schl{\"u}ter}
\affil[1]{Universidad Aut\'onoma de Guadalajara}
\affil[2]{Ulm University}
\affil[3]{Ulm University of Applied Sciences}

\date{}

\begin{document}

\maketitle

\abstract{Hyperlocal meteorological sensing is essential for accurate
solar photovoltaic forecasting, yet professional-grade meteorological
stations require investments easily exceeding \$1{,}000~USD per node, making
distributed deployments economically inaccessible.
This work presents a modular internet of things (IoT) device based on the ESP32
microcontroller integrating temperature, humidity, luminosity, and solar
irradiance sensors in an IP68-rated enclosure at a total hardware costs
of about \$65~USD when components are sourced in Germany.
A hybrid architecture decouples external model training, performed on a
conventional computer using the software Python and the open-source library TensorFlow, from autonomous
24-hour solar voltage forecasting executed on-device via a three-layer
feedforward network with 3{,}011 parameters (11.8\,KB). The network is trained offline
on site-collected data and deployed on the microcontroller as static
weight matrices without cloud connectivity.
An on-device incremental gradient descent mechanism enables continuous
model adaptation after deployment without external retraining.
The system was evaluated through two field deployments: a short period of
hardware and firmware validation in Ulm, Germany, and a 115-day deployment
in Zapopan, Mexico, comprising 84~days of training and 31~days of autonomous
operation with zero missing records.
Over a clean 28-day daytime window, the embedded model attained a coefficient
of determination of 0.9165 and a mean absolute error of 0.2975~V (4.65\% of the
operational range), outperforming a climatology baseline (skill score 0.64)
while not surpassing a 24-hour persistence baseline.
A frozen-weight ablation confirms that the on-device update mechanism yields a
small but statistically robust accuracy gain ($p = 0.001$), demonstrating that
autonomous incremental learning is feasible on low-cost hardware without cloud
connectivity.
}
\vfill
\textbf{keywords:}{Low-cost sensors; Forecasting; Feedforward neural network; ESP32 microcontroller; Modular design; embedded inference; TinyML}

\section{Introduction}

Accurate short-horizon forecasting of photovoltaic (PV) generation critically depends on hyperlocal meteorological measurements.
Irradiance, temperature, and humidity vary significantly across distances
of only a few hundred meters, and site-level variations of this magnitude
are invisible to both national observation networks, which deploy at most
one station per urban area, and satellite reanalysis products such as
NASA MERRA-2 \cite{Merra2}, which operate at spatial resolutions of
approximately 50~km.
The result is a persistent data gap: the measurements needed to drive
practical PV management are available neither from sparse ground networks
nor from coarse-resolution remote sensing, yet professional-grade
meteorological stations capable of filling this gap require investments easily
exceeding \$1{,}000~USD per node \cite{caine2015mobile, buildings11080336}.

Embedded Internet of Things (IoT) platforms offer a structural solution to this gap.
Several authors have deployed low-cost microcontroller-based systems for
environmental monitoring. In \cite{samara2019intelligent}, authors demonstrated
artificial neural network (ANN)-based PV prediction on a microcontroller with high
accuracy, whereas in \cite{SRIDHAR2023100609}, authors proposed a modular ESP32 platform
for on-device air-quality inference. Similarly in \cite{Melo221} authors implemented
real-time PV monitoring with local data storage.
However, none of these systems combines autonomous solar energy
forecasting, continuous post-deployment model adaptation, and
operation without cloud connectivity in a single embedded device.
The deployment of computationally demanding, i.e.\ energy-intensive, neural
networks is one of the crucial problems. As a matter of fact, even the
integration of comparably simple recurrent neural network (RNN) architectures
on resource-constrained microcontrollers remains an open challenge
\cite{Ahmed2022}.

We address this gap with a modular IoT device that resolves the
fundamental tension between algorithmic sophistication and embedded
hardware constraints through a hybrid processing architecture.
A three-layer feedforward network is trained offline in Python and
TensorFlow on site-collected environmental data; the resulting weight
matrices are then deployed on an
ESP32 microcontroller \cite{Kareem2021} as static arrays for fully
autonomous 24-hour solar voltage forecasting without cloud connectivity.
An on-device incremental gradient descent mechanism
\cite{Losing2018incremental} allows the deployed model to adapt to local
conditions without external retraining.
The assembled hardware integrates a DHT22 sensor \cite{puspasari2020accuracy},
an LDR luminosity sensor, a DS3231 real-time clock \cite{Putri23}, and a
miniature solar panel as irradiance proxy, all housed in an IP68 enclosure,
at a total cost of about \$65~USD\footnote{Construction cost based on local
component sourcing. Appendix~\ref{tab:costs} lists individual retail unit
prices as an upper reference (German online retail prices, July~2026); costs are indicative
and subject to supplier and regional variation.}, communicating via Bluetooth Low Energy
with a companion mobile application \cite{patton2019app, Erick2024}.
The system is evaluated through two field deployments.
A short period of testing in Ulm, Germany served as hardware and firmware
validation, with measurements cross-checked against a weather station operated by the German Climate Service and
MERRA-2 reference data \cite{Merra2}.
A subsequent 115-day deployment in Zapopan, Mexico (84~days training,
31~days autonomous validation, zero missing records) provided the
quantitative evaluation of the forecasting pipeline reported in
Section~\ref{sec:Case Study}.

The remainder of this article is structured as follows.
Section~\ref{sec:hardware} describes the hardware design.
Section~\ref{sec:softwareDesign} presents the forecasting algorithm
and its embedded implementation.
Section~\ref{sec:Case Study} reports both case studies.
Section~\ref{sec:discussion} contextualizes the results against related
work, and Section~\ref{sec:conclusion} presents conclusions and future
directions.

\section{Hardware}
\label{sec:hardware}

The hardware design is guided by three primary requirements: robustness for
outdoor field deployment (particularly waterproofing), cost-efficiency to
enable wide accessibility, and modularity to support future sensor expansions.
Table~\ref{tab:bom} lists the eight commercially available off-the-shelf
components that constitute the system. Their combined cost does not exceed
\$55~USD when sourced locally in Mexico, representing a reduction of more than 99\% compared to professional
meteorological stations that typically require investments exceeding
\$1{,}000~USD per node \cite{buildings11080336, caine2015mobile}. In Appendix~\ref{tab:costs} we give a full cost breakdown including exemplary supplier.

\begin{table}[ht]
\caption{System components and their functions}
\label{tab:bom}
\begin{tabularx}{\textwidth}{llX}
\toprule
\textbf{Component} & \textbf{Model} & \textbf{Function} \\
\midrule
Microcontroller       & ESP32 WROOM32           & Central processing, bluetooth low energy
                                                  communication, data logging \\
Temp./Humidity sensor & DHT22                   & Temperature \& relative
                                                  humidity measurement \\
Luminosity sensor     & LDR (GL5528)            & Ambient light intensity
                                                  measurement \\
Real-time clock       & DS3231                  & Timestamp synchronization
                                                  for time-series integrity \\
Solar panel           & 9\,V PV (adj.\ 6.4\,V) & Localized irradiance proxy
                                                  via voltage output \\
MicroSD module        & SPI interface           & Local JSON data storage \\
IP68 enclosure        & Weatherproof ABS        & Dust and water ingress
                                                  protection for field deployment \\
Printed circuit board \& connectors     & Custom modular design   & Integrated wiring with
                                                  expansion ports \\
\bottomrule
\end{tabularx}
\end{table}

\subsection{Development Platform: ESP32 WROOM32}
\label{sec:esp32}

The ESP32 WROOM32 module was selected as the central processing unit of the device. Its dual-core Xtensa LX6 processor operating at up to 240\,MHz
provides the computational capacity required for real-time feedforward
inference, while its deep-sleep current draw of only 10\,$\mu$A enables extended battery
operation in field deployments \cite{Kareem2021, babiuch2019using}. Key
specifications relevant to this application include 520\,KB SRAM and 4\,MB
flash memory for local model storage, a 12-bit ADC for direct sensor
interfacing, and integrated Wi-Fi~802.11~b/g/n and Bluetooth~4.2/BLE for
wireless data transmission \cite{Nkemeni2020}. In this project, the ESP32
coordinates the acquisition of temperature, humidity, luminosity, and solar
panel voltage, generates real-time solar voltage forecasts, stores data on a microSD
card in JSON format, and transmits results to a mobile application via
Bluetooth Low Energy~(BLE).

\subsection{Sensors and Measurement Technologies}
\label{sec:sensors}
Sensor selection was guided by four criteria: measurement accuracy, operational reliability across diverse climates,
energy efficiency, and compatibility with the ESP32 platform. The resulting
sensor suite covers all environmental variables required for solar energy
forecasting: temperature, humidity, luminosity, and irradiance proxy via a
small solar panel.

\paragraph{(a) DHT22 Temperature and Humidity Sensor.}
The DHT22 sensor provides temperature measurements over a range of
$-$40\,$^{\circ}$C to $+$80\,$^{\circ}$C with an accuracy of
$\pm$0.5\,$^{\circ}$C, and relative humidity (RH) measurements from 0 to 100\%
with an accuracy of $\pm$2--5\%~RH \cite{puspasari2020accuracy,Wardani2023}.
Its digital one-wire communication protocol minimizes reading errors compared
to analog sensors, providing greater measurement stability under variable
environmental conditions. The DHT22 was selected over alternative sensors
(e.g., BME280) because its temperature range fully covers both deployment
sites, Ulm, Germany (recorded range: 10.2--35.3\,$^{\circ}$C) and Zapopan,
Mexico (recorded range: 8.2--53.0\,$^{\circ}$C), at 40--60\% lower initial
cost \cite{Sunardi2023}. It should be noted that the sensor is mounted inside
the enclosure, so daytime readings include a radiative heating bias relative to
ambient air temperature: values above 45\,$^{\circ}$C occur exclusively between
11:00 and 17:00 local time, whereas nocturnal maxima remain below
23\,$^{\circ}$C. The recorded temperature is therefore interpreted throughout
this work as an enclosure temperature that co-varies with ambient conditions,
not as a calibrated air-temperature measurement.

\paragraph{(b) DS3231 Real-Time Clock Module}
Temporal synchronization is provided by the DS3231 real-time clock~(RTC)
module, which supplies accurate timestamps essential for maintaining the
integrity of the time-series data and for the sliding-window input structures
used by the forecasting model \cite{Putri23}. Its temperature-compensated crystal
oscillator ensures clock accuracy across the full operating temperature range
encountered at both deployment sites.

\paragraph{(c) Light-Dependent Resistor Luminosity Sensor}
A light-dependent resistor (LDR, model GL5528) measures ambient light
intensity by exploiting the inverse relationship between incident light and
electrical resistance. Its detection range of 10 to 100{,}000\,lux covers
conditions from dawn to peak solar irradiance \cite{Wu2011/robio.2011.6181275}.
The GL5528 exhibits peak spectral sensitivity near 540\,nm (visible spectrum),
making it suitable as a relative irradiance proxy under clear-sky and partially
cloudy conditions; resistance-to-lux conversion follows the manufacturer's
characteristic curve applied during firmware initialization.

\paragraph{(d) Solar Panel as Irradiance Sensor}
A 9\,V photovoltaic panel (nominal power 1\,W; open-circuit voltage 9\,V at
1000\,W/m$^2$, 25\,$^{\circ}$C) serves as a localized irradiance sensing
element. Its output is conditioned through a resistive voltage divider so that
the full-scale signal matches the ADC input range of the ESP32, giving an
operational span of 0--6.4\,V. The panel voltage is a monotonic but non-linear
function of incident irradiance, and it is additionally affected by cell
temperature; it is therefore used as an uncalibrated relative proxy rather than
as a quantitative irradiance measurement in W/m$^2$. Sampled every 15
minutes \cite{wan2015photovoltaic}, this signal captures microclimate
variations, including precipitation events detectable as characteristic
voltage drops during theoretically optimal solar conditions, that are
unavailable from satellite reanalysis products at 50\,km resolution.

\subsection{Hardware Integration and Enclosure}
\label{sec:HardwareSetup}

\begin{figure}[ht!]
    \begin{center}
        \vspace{5mm}
        \includegraphics[scale=0.30]{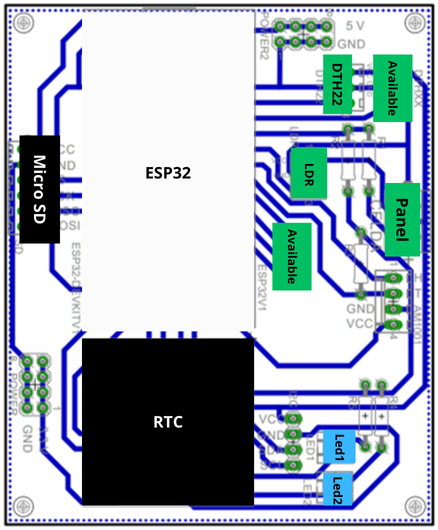}
        \caption{PCB design of the monitoring and forecasting system,
                 showing the spatial layout of the ESP32 microcontroller,
                 RTC module, MicroSD module, and sensor connection ports.
                 Labeled expansion slots allow future sensor additions
                 without redesign.}
        \label{fig:pcb}
    \end{center}
\end{figure}

All components are integrated onto a custom printed circuit board (PCB) whose layout reflects the modular architecture of the system.
Figure~\ref{fig:pcb} illustrates the PCB design, showing the spatial
arrangement of all modules and the interconnection paths between them.
The ESP32 is centrally located on the PCB, acting as the system's processing hub. The RTC module is positioned directly below the ESP32 to minimize the I$^2$C bus length and reduce timing jitter. The MicroSD module is placed on the left side for convenient card access without disassembly. The DHT22 and LDR sensors are located in the upper-right section, physically isolated from the heat-generating microcontroller to prevent thermal cross-interference with temperature readings. Two labeled expansion ports allow future integration of additional digital or analog sensors without PCB redesign. Male and female connector headers allow individual sensor modules to be replaced or
repositioned without soldering, and data are stored in JSON format on the
microSD card to facilitate direct import into the external Python training pipeline described in Section~\ref{sec:softwareDesign}.

\begin{figure}[ht]
    \centering
    \begin{subfigure}{0.45\linewidth}
        \includegraphics[scale=0.50]{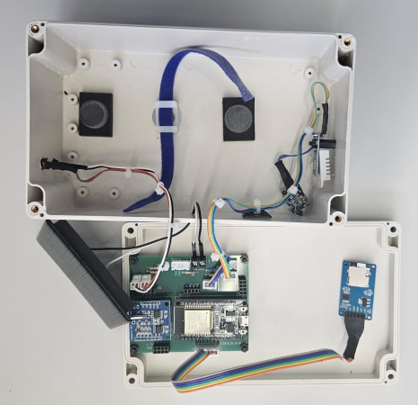}
        \caption{Internal component arrangement showing the ESP32,
                 sensor modules, RTC, and MicroSD card mounted on
                 the custom PCB with modular connectors.}
        \label{fig:construccion}
    \end{subfigure}
    \hfill
    \begin{subfigure}{0.45\linewidth}
        \includegraphics[scale=0.05]{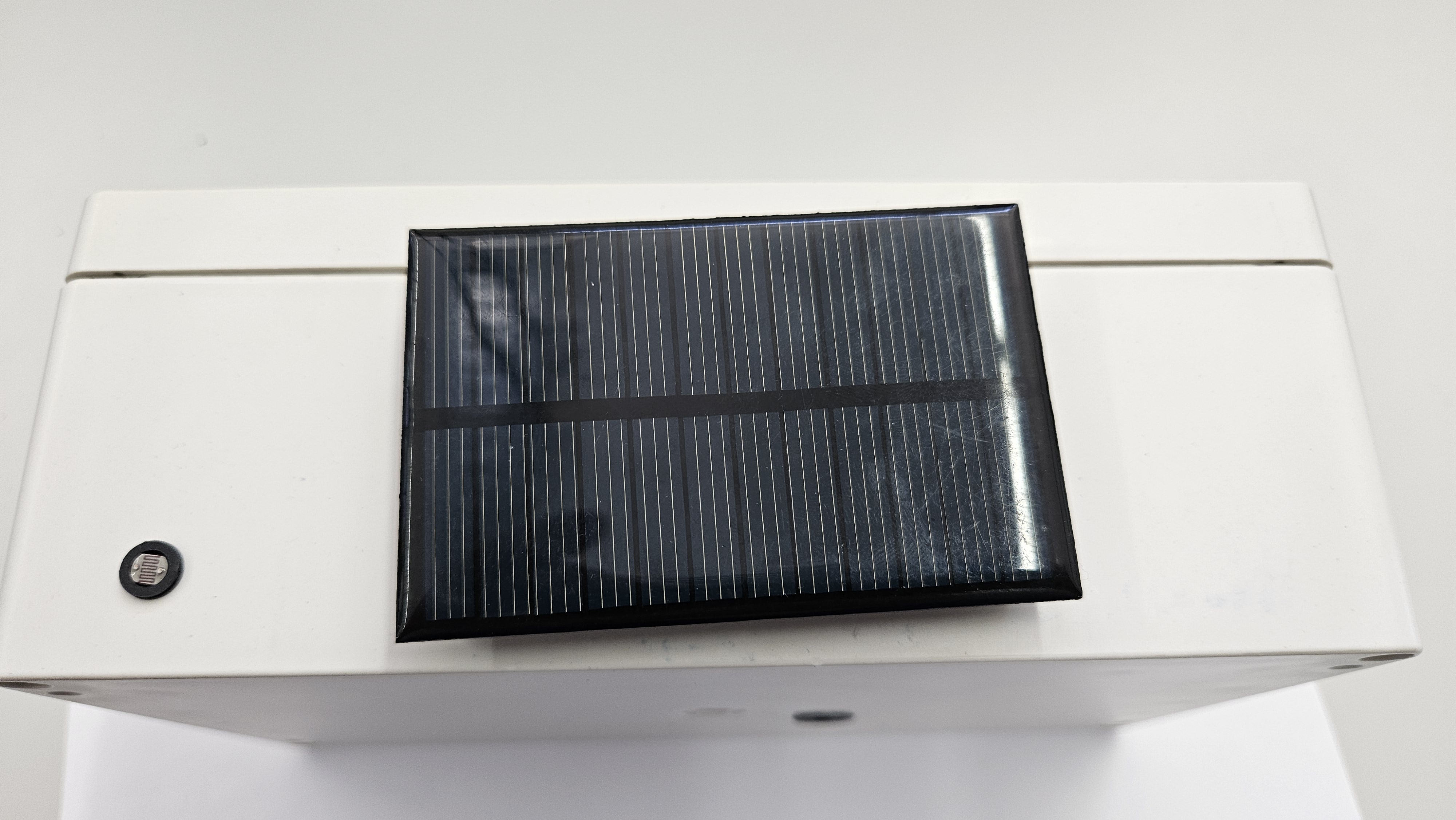}
        \caption{Sealed IP68 enclosure with the solar panel and LDR
                 sensor mounted on the exterior face for unobstructed
                 irradiance measurement.}
        \label{fig:closed}
    \end{subfigure}
    \caption{Assembly of the monitoring and forecasting system with ESP32:
             internal layout (left) and field-deployed enclosure (right).}
    \label{fig:construccion_all}
\end{figure}

Environmental protection is provided by an IP68-rated enclosure, which
guarantees complete dust ingress protection and resistance to continuous
water immersion beyond one metre depth \cite{Abdelmoneim2023}. This
classification is essential for unattended outdoor deployment, where
prolonged exposure to rain, humidity, and particulate matter can compromise
electronic components and introduce systematic measurement errors. The
enclosure was validated across both case study environments: the subtropical
conditions of Zapopan, Mexico (relative humidity reaching 90\% during the
rainy season) and the late-spring conditions of Ulm, Germany.
Figure~\ref{fig:construccion_all} shows the assembled device, including the
internal component arrangement (Figure~\ref{fig:construccion}) and the
sealed field-ready enclosure with the solar panel and LDR mounted on the
exterior face for unobstructed sky exposure
(Figure~\ref{fig:closed}).

\section{Software}
\label{sec:software}

The software needs to satisfy multiple requirements: it has to be lean in order to run on the ESP32 microcontroller, it has to control the sensors described in Section \ref{sec:hardware} for recording and storing the data and it has to produce forecasts. For producing forecasts, we use a feedforward neural network as motivated in Section \ref{ffn}. As training a neural network is computationally demanding we split the whole process into an offline phase (done externally) and an online phase (done on the device). The network is trained offline using site-collected data, and then updated on the sensor box using live data. Together, these two stages constitute the hybrid architecture that resolves the fundamental tension between algorithmic sophistication and the computational constraints of low-cost IoT hardware.%

\subsection{Feedforward Networks}
\label{ffn}

Neural networks are an established and widely used tool for data analytics; hence we do not give a detailed introduction and refer, e.g., to \cite{aggarwal2018neural} instead. For time series forecasting, so-called long short-term memory (LSTM) neural networks \cite{hochreiter1997long} and convolutional neural networks have both been used \cite{al2020comprehensive, kreuzer2020short}; however, their training is computationally demanding and cannot be repeated every time new data arrive. Instead, the literature suggests a regular parameter update, say every week \cite{Losing2018incremental, LIU2019392}. Alternatively, for a more lean handling we might use simple statistical methods such as the autoregressive integrated moving average (ARIMA) model. Its implementation is very efficient and can be easily handled by our chosen processor; however, multiple authors such as \cite{kreuzer2020short} have shown that it is clearly outperformed by neural networks. Hence, to find a balance between complexity and precision, we opt for a feedforward neural network. This type of network processes information only in one direction without any feedback loops -- contrary to recurrent neural networks such as LSTM networks. Once the network is trained as explained in Section~\ref{sec:annDesign}, this allows us to use a simplified parameter update mechanism which is explained in Section \ref{sec:sliding}.

\subsection{External Training of the Embedded Model}
\label{sec:annDesign}

The forecasting model deployed on the ESP32 is a feedforward network with an architecture that reflects our choice for the sliding window explained in Section~\ref{sec:sliding}: here we consider quarter-hourly measurements and the past 96 observations in our model. We opted for this length, as the window spans exactly 24 hours of measurements, which is the periodicity of the solar cycle and the natural horizon for diurnal forecasting \cite{Bhatt2022sliding}. However, alternative window sizes are possible, of course. This results in an 288/10/10/1 neural network setup: an input layer of 288 values
(past 96 measurements of three input features), two hidden layers of 10 neurons each
with hyperbolic tangent activation, and a single linear output
neuron. Training is performed offline on a conventional computer using Python
with TensorFlow and Keras \cite{tensorflow2015}, with pandas and NumPy
handling data manipulation and matrix operations. The three
input features are temperature ($^\circ$C), relative humidity (\%), and
luminosity (expressed as a 0--100 relative index from the LDR, not calibrated lux); the target output is solar panel voltage~(V), which serves as an uncalibrated proxy for local irradiance.

Raw sensor readings are preprocessed to remove inconsistencies and fill
missing values, then normalized using min-max normalization with sensor-specific bounds computed from the training set and stored as firmware constants, ensuring that inference-time scaling is identical to training-time scaling regardless of observed data range \cite{meenal2022weather, Simeaunu2020sliding}. The temporal structure of the training data is constructed using the sliding window formalism defined in Section~\ref{sec:sliding} below. The network is compiled with the
Adam optimizer \cite{kingma2014adam} and mean squared error (MSE) as the loss
function. Training runs for up to 200 epochs with an 80/20 train--validation
split; early stopping halts training when validation loss does not improve for
10 consecutive epochs \cite{thakur2024}. Once training converges, the optimized
weight matrices and bias vectors are extracted and transferred to the ESP32 as
static \texttt{float32} arrays stored in flash memory.

\subsubsection{Sliding Window Mechanism}
\label{sec:sliding}

The sliding window is the central data structure that links the
training pipeline, the on-device inference engine, and the adaptive
update mechanism \cite{Simeaunu2020sliding, Bhatt2022sliding,
LIU2019392}. At discrete time step $t$, the active window
$\mathcal{W}_t$ consists of the 96 most recent consecutive
measurements:

\begin{equation}
    \mathcal{W}_t = \bigl(
        \mathbf{x}_{t-95},\;
        \mathbf{x}_{t-94},\;
        \ldots,\;
        \mathbf{x}_{t}
    \bigr)
    \label{eq:window_def}
\end{equation}

where each $\mathbf{x}_\tau \in \mathbb{R}^{3}$ is the vector of
sensor readings (temperature, humidity, luminosity) at the 
15-minute interval indexed by $\tau$. Once a new measurement $\mathbf{x}_{t+1}$ becomes available, the window advances by discarding the oldest observation and incorporating the new one:

\begin{equation}
    \mathcal{W}_{t+1}
        = \bigl(
            \mathbf{x}_{t-94},\;
            \ldots,\;
            \mathbf{x}_{t},\;
            \mathbf{x}_{t+1}
          \bigr)
    \label{eq:window_update}
\end{equation}

This operation is implemented as a circular buffer of
96 positions stored in JSON format on the microSD card \cite{Erick2024}. When the write pointer reaches position 97, it wraps to position 1 and overwrites the oldest entry, requiring only a single write operation per acquisition cycle with no memory reallocation \cite{Simeaunu2020sliding}. The sliding window mechanism avoids computational and memory overhead of storing
the full historical dataset on the device while ensuring that
inference and weight updates always reflect the most recent
environmental conditions \cite{Losing2018incremental}.

\subsubsection{Embedded Model Deployment and Inference Pipeline}
\label{sec:embedded}

The model deployed on the ESP32 is identical to the external model specified in Section \ref{sec:annDesign}. The input to the network is a flattened
vector of 288 values derived from the 96-sample sliding window
(96 measurements $\times$ 3 features: temperature, humidity, and luminosity).

This architecture yields a total of 3{,}011
trainable parameters, which at \texttt{float32} precision occupy
11.8\,KB of the ESP32's 4\,MB flash memory, leaving ample flash for
firmware and the SRAM free for the circular buffer and BLE
communication \cite{Kareem2021, babiuch2019using}. The inference
pipeline follows four steps executed sequentially on the device at each
prediction cycle:

\paragraph{Step 1: Input normalization (Eq.~\ref{eq:norm}).}
When the circular buffer $\mathcal{W}_{t+1}$ contains 96 samples,
a prediction cycle is triggered. Each element $x$ of each input
vector is normalized to $[0,\,1]$ using sensor-specific bounds
established during the training phase:

\begin{equation}
    x_{\mathrm{norm}}
        = \frac{x - x_{\min}}{x_{\max} - x_{\min}}
    \label{eq:norm}
\end{equation}

Min-max normalization (Eq.~\ref{eq:norm}) is the standard
preprocessing choice for neural networks operating on heterogeneous
physical sensors, as it eliminates inter-variable scale biases
without assumptions about the underlying
distribution \cite{Simeaunu2020sliding, meenal2022weather}. The
bounds $x_{\min}$ and $x_{\max}$ are stored as constants in firmware
\cite{Erick2024}, guaranteeing that inference-time scaling is
identical to training-time scaling regardless of observed data range.

\paragraph{Step 2: Forward propagation with tanh activation
(Eq.~\ref{eq:forward}).}
The normalized 96-point input sequence is propagated through the
two hidden layers of the embedded network. For each layer $l$
($l = 1,\,2$) we compute:

\begin{equation}
    \mathbf{h}^{(l)}
        = \tanh\!\left(
            \mathbf{W}^{(l)}\,\mathbf{h}^{(l-1)}
            + \mathbf{b}^{(l)}
          \right)
    \label{eq:forward}
\end{equation}

where $\mathbf{W}^{(l)}$ and $\mathbf{b}^{(l)}$ are the static
weight matrix and bias vector of layer $l$ stored in flash memory,
and $\mathbf{h}^{(0)}$ is the normalized input. The hyperbolic
tangent activation function was selected because it preserves the
sign of neuron activations, enabling the network to model both the
positive rising slope and the plateau-and-fall structure of the
diurnal voltage curve \cite{Bhatt2022sliding}. The forward pass
is implemented entirely via matrix–vector multiply-accumulate
operations, which are natively supported by the ESP32's Xtensa LX6
cores without requiring floating-point co-processor
support \cite{babiuch2019using, Erick2024}. A special case is
enforced in firmware: when the measured luminosity is zero
(nighttime), the output is set directly to $0$\,V, overriding the
network result and preventing physically impossible non-zero
nocturnal predictions \cite{Erick2024}.

\paragraph{Step 3: Output denormalization and saturation
(Eq.~\ref{eq:denorm}).}
The raw scalar output $\hat{y}_{\mathrm{norm}}$ of the final layer
is converted to physical voltage units:

\begin{equation}
    \hat{v}
        = \hat{y}_{\mathrm{norm}}\,
          \bigl(v_{\max} - v_{\min}\bigr) + v_{\min}
    \label{eq:denorm}
\end{equation}

where $v_{\min} = 0$\,V and $v_{\max} = 6.4$\,V are the
operational limits of the solar panel after voltage division
(Section~\ref{sec:hardware}). Hard saturation clamps
$\hat{v} \in [0,\,6.4]$\,V, suppressing physically impossible
predictions under atypical input conditions \cite{Erick2024}.
The denormalization inverse of Eq.~\ref{eq:norm} is standard
practice in embedded neural network deployment to recover
meaningful physical outputs from normalized model
predictions \cite{Simeaunu2020sliding}.

\paragraph{Step 4: On-device adaptive weight update
(Eq.~\ref{eq:update}).}
After each completed 24-hour forecast day, the static weight
matrices and bias vectors are updated in-place using a simplified gradient descent
step applied to the current window $\mathcal{W}_{t+1}$
(Eq.~\ref{eq:window_update}):

\begin{equation}
    \mathbf{W}^{(l)}
        \leftarrow
        \mathbf{W}^{(l)}
        - \eta\;
          \nabla_{\mathbf{W}^{(l)}}
          \mathcal{L}\!\left(
              \hat{\mathbf{v}},\;\mathbf{v}
          \right)
    \label{eq:update}
\end{equation}

where $\mathcal{L}$ is the mean squared error between the previous
day's predicted voltage sequence $\hat{\mathbf{v}}$ and the
corresponding measured values $\mathbf{v}$ drawn from
$\mathcal{W}_{t+1}$. Equation~\ref{eq:update} is written for the weight
matrices; the bias vectors are updated by the analogous rule
$\mathbf{b}^{(l)} \leftarrow \mathbf{b}^{(l)} - \eta\,\nabla_{\mathbf{b}^{(l)}}\mathcal{L}$.
The parameter $\eta = 0.001$ is a fixed learning rate chosen small enough to avoid destabilizing the trained weights while allowing gradual adaptation to local conditions. This value was selected empirically during offline experimentation; the energy cost of the update step was not instrumented on the deployed device. The
gradient $\nabla_{\mathbf{W}^{(l)}}\mathcal{L}$ is computed via
standard backpropagation through the embedded
layers \cite{werbos1990backpropagation, alqushaibi2020review}. Critically, this update
operates exclusively on the sliding window (Eq.~\ref{eq:window_def})
where no historical data outside the current 96-point buffer are
required, and no connection to an external server is
needed \cite{Losing2018incremental}. This design choice directly
addresses a key limitation identified in the incremental learning
literature: the impracticality of maintaining permanent cloud
connectivity for real-time model updates in resource-constrained
IoT deployments \cite{Losing2018incremental}. The complete update
procedure is implemented in firmware and verified in a publicly
available source code \cite{Erick2024}.

The key property of this mechanism is that the model adapts to
local seasonal drift and site-specific micro-climate patterns
without requiring cloud connectivity, external retraining, or
physical access to the device \cite{Losing2018incremental}. Because
the update operates exclusively on the 96-point sliding window
(Eq.~\ref{eq:window_update}), memory usage remains constant
regardless of deployment duration, which is a critical requirement
for long-term autonomous operation on memory-constrained
hardware \cite{Losing2018incremental, Erick2024}. The empirical
performance of this adaptive mechanism across both case study
deployments is quantified and discussed in
Section~\ref{sec:Case Study}.

\subsection{The Software Design on the Sensor}
\label{sec:softwareDesign}

The ESP32 firmware coordinates the four functional layers described
above within a continuous event-driven loop. Every 15 minutes the
device reads temperature and relative humidity from the DHT22,
light intensity from the LDR GL5528, and solar panel voltage via
the 12-bit ADC; the DS3231 RTC provides a synchronized timestamp
for each measurement. All readings pass through basic validity
filters before being appended as a JSON record to the circular
buffer on the microSD card, advancing the sliding window
$\mathcal{W}_t$ by one position (Eq.~\ref{eq:window_update}).
Between acquisition cycles the microcontroller enters deep-sleep
mode ($\leq 10\,\mu$A), significantly extending operational
lifetime when powered by a power
bank \cite{babiuch2019using}.

Once 96 timestamped samples are available, the device executes the
four-step inference pipeline (Eqs.~\ref{eq:norm}–\ref{eq:denorm}),
generating a 24-hour solar voltage forecast. The predicted sequence
is stored locally alongside the measured data and made available for
BLE transmission on demand. Following each completed forecast day,
the weight update (Eq.~\ref{eq:update}) is applied in-place before
the buffer advances, implementing on-device adaptation
without external intervention \cite{Losing2018incremental,
Erick2024}. The modular software architecture ensures that each
functional layer, namely acquisition, storage, prediction, and
communication, operates independently with well-defined interfaces,
simplifying maintenance and facilitating future integration of
alternative communication protocols such as LoRa or
Wi-Fi \cite{Nkemeni2020}.

\subsection{Mobile Application}

\begin{figure}[ht!]
    \centering
    \begin{subfigure}{0.3\linewidth}
        \includegraphics[width=\linewidth]{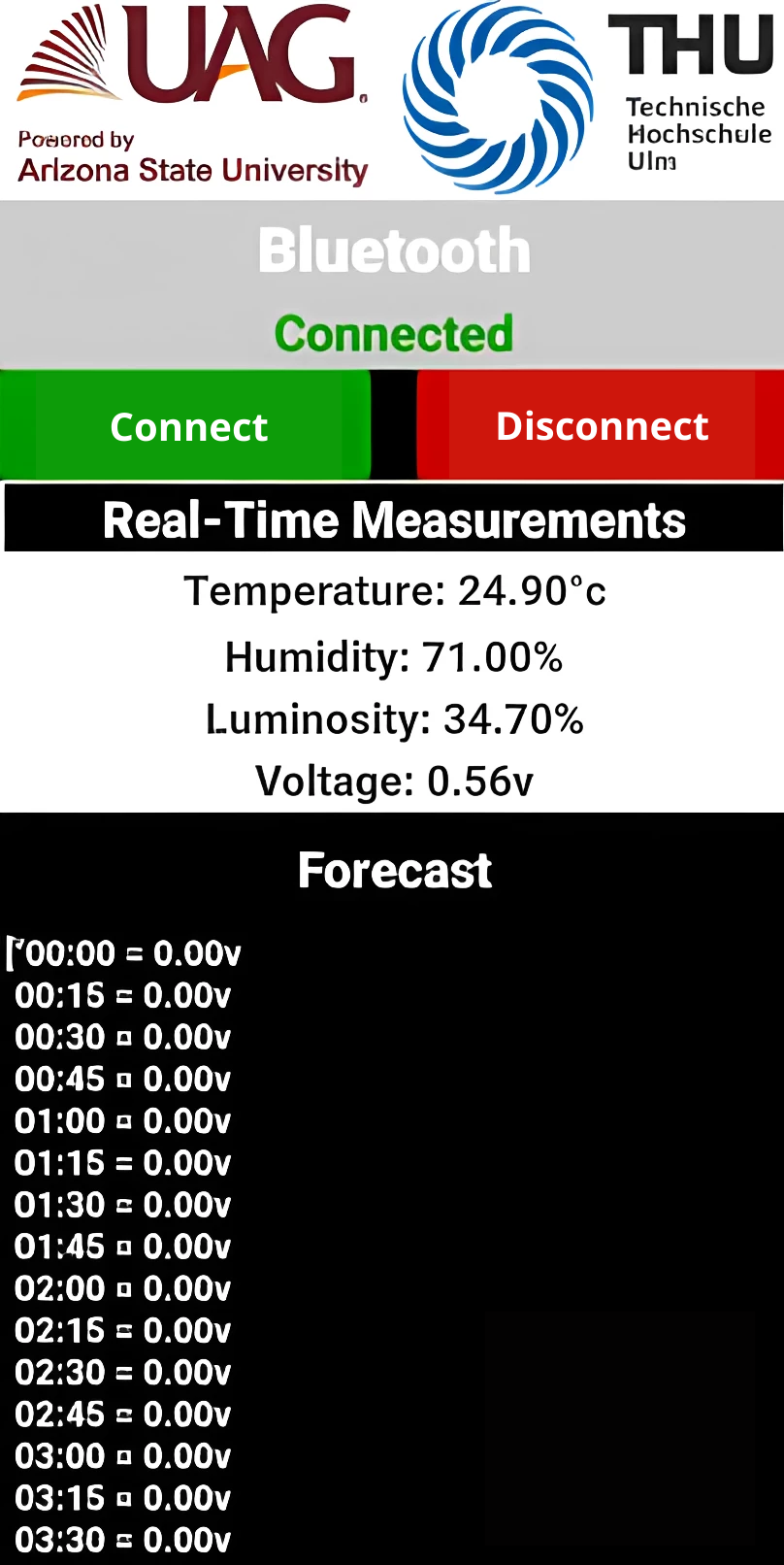}
        \caption{Interface in Connected State.}
        \label{fig:Connected}
    \end{subfigure}
    \hspace{0.05\linewidth}
    \begin{subfigure}{0.3\linewidth}
        \includegraphics[width=\linewidth]{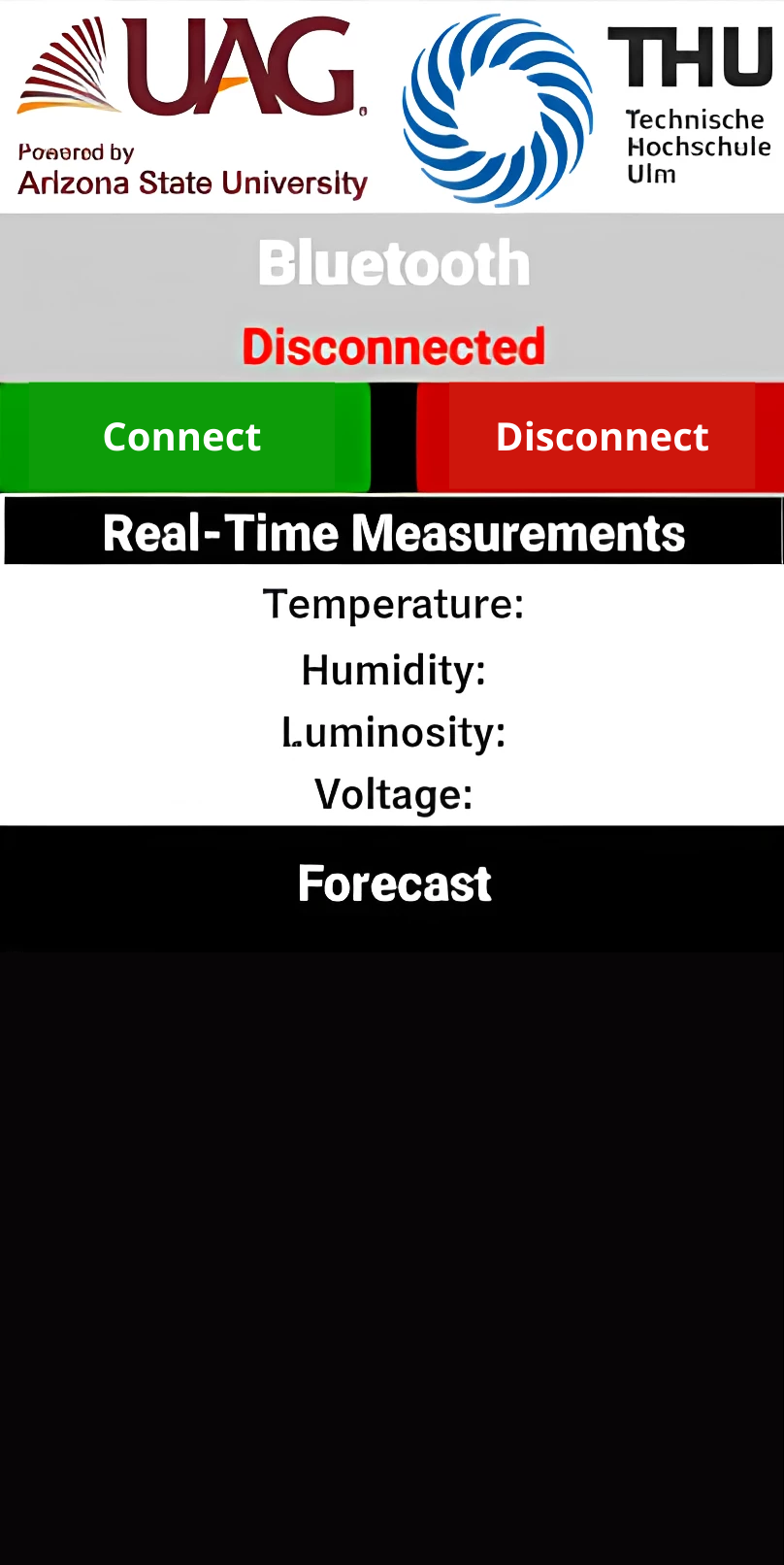}
        \caption{Interface in Disconnected State.}
        \label{fig:Disconnected}
    \end{subfigure}
    \caption{Screenshots of the mobile application interface
             developed with MIT App Inventor \cite{patton2019app}.
             The connected state (left) displays real-time sensor
             readings and the 24-hour solar voltage forecast
             generated by the embedded feedforward network; the
             disconnected state (right) allows Bluetooth device
             selection.}
    \label{fig:MobileApp}
\end{figure}

The mobile application is developed using the MIT App Inventor
platform \cite{patton2019app}, providing an intuitive interface for
monitoring real-time environmental data and displaying the 24-hour
solar energy forecasts generated by the
device \cite{Erick2024}.
Figure~\ref{fig:MobileApp} shows the two operational states:
a connected state providing live sensor readings and forecasts
(Figure~\ref{fig:Connected}), and a disconnected state for device
selection (Figure~\ref{fig:Disconnected}).

Once a device is selected, the application connects automatically
and updates its status. A background timer periodically queries the
sensor for the latest measurements; when data are available they
are parsed and the forecast display is refreshed in real time. Each
connection cycle triggers a request for the predicted solar voltage
for the upcoming 24 hours, which is the same forecast window defined by
the sliding window $\mathcal{W}_t$ (Eq.~\ref{eq:window_def}) on
the device. The application includes mechanisms to handle connection
interruptions and supports offline mode, allowing users to review
previously received data when the sensor is not actively paired.
The full source code is publicly available to encourage community
adaptation and extension \cite{Erick2024}.

\section{Case Study}
\label{sec:Case Study}

The system was evaluated through two sequential field
deployments carried out in different geographic and climatic
environments. The first deployment took place in Ulm, Germany,
and served as an initial operational test of the complete
system. The second deployment was conducted in Zapopan,
Mexico, with an extended data collection and training period
that enabled a more comprehensive quantitative evaluation of
the forecasting pipeline.

\subsection{Case Study I: Ulm, Germany}

\subsubsection{Experimental Setup and Location}

The first deployment was carried out on the rooftop of a residential building in Ulm, Germany, from May 26 to June 14,
2024. The site was selected for two main reasons: the
availability of public reference meteorological data from the
German Climate Service (DWD, \textit{Deutscher Wetterdienst},
\texttt{www.dwd.de}), and the proximity to the Ulm University
of Applied Sciences campus where the prototype was constructed
and calibrated, enabling systematic fault detection before
extended autonomous operation.

\begin{figure}[ht]
    \centering
    \includegraphics[width=0.6\linewidth]{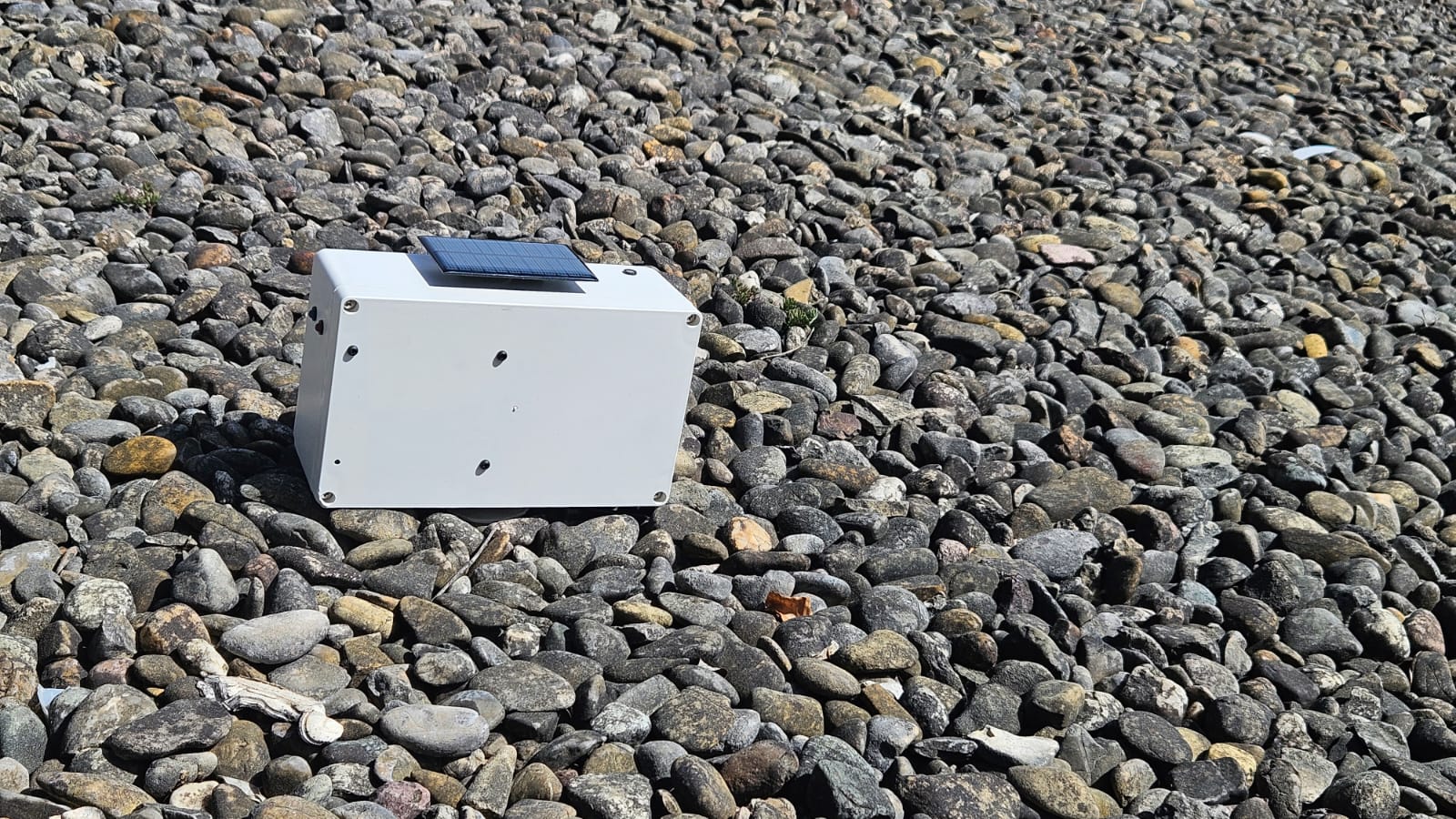}
    \caption{The assembled device deployed on the rooftop site in
             Ulm, Germany, with the solar panel and LDR exposed for
             unobstructed sky exposure.}
    \label{fig:ulm_deploy}
\end{figure}

Environmental parameters (temperature, relative humidity,
and solar panel voltage) were recorded at 15-minute
intervals. Measurements were compared against two
independent reference sources: the DWD local weather station
and the NASA MERRA-2 reanalysis \cite{Merra2}, which provides
gridded atmospheric data at $0.5^\circ \times 0.625^\circ$
spatial resolution. Figure~\ref{fig:ulm_deploy} shows the
device installed at the Ulm rooftop site.

\subsubsection{Recorded Environmental Data}

\begin{figure}[H]
    \centering
    \includegraphics[scale=0.30]{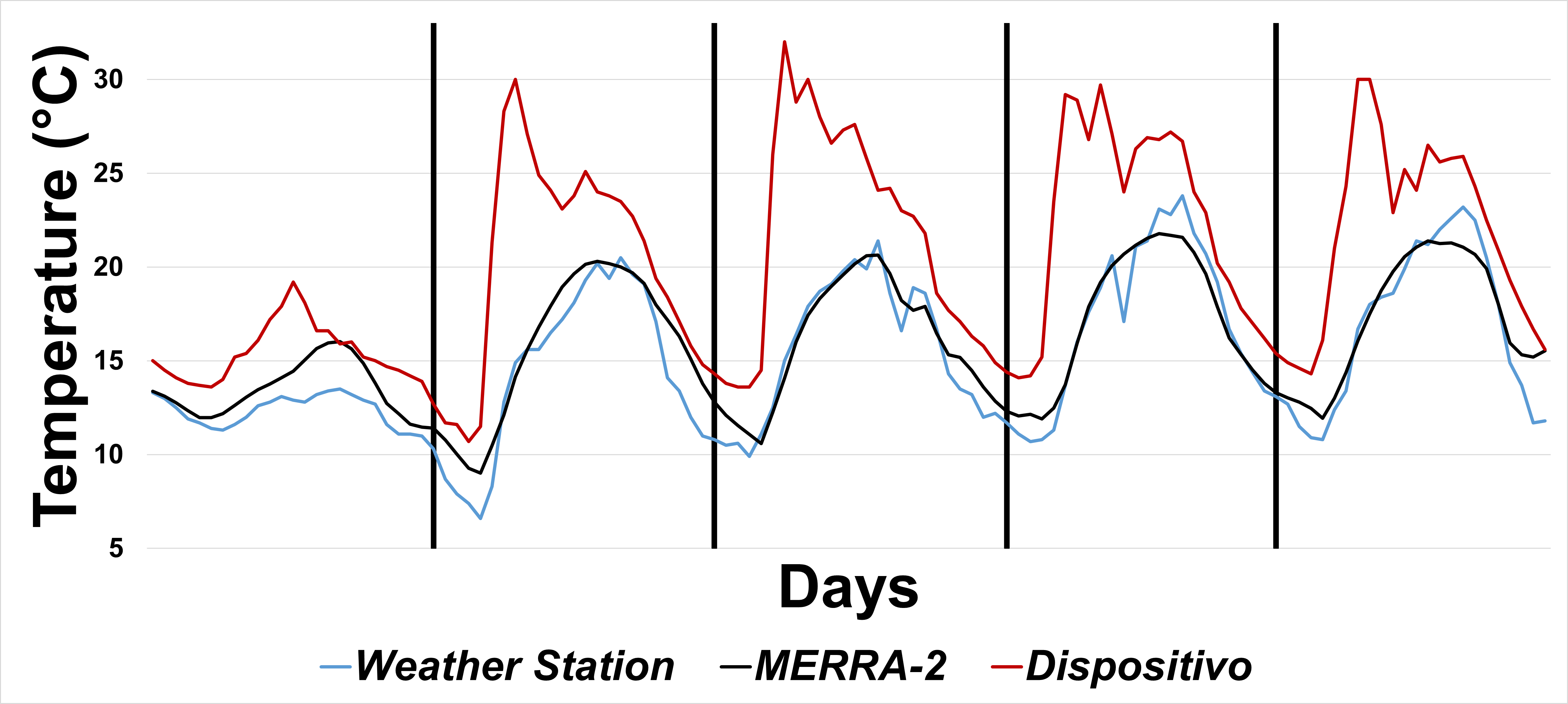}
    \caption{Temperature recorded by the device during the
             Ulm deployment, compared against DWD station
             and MERRA-2 reanalysis reference data.}
    \label{fig:Temperature}
\end{figure}

\begin{figure}[H]
    \centering
    \includegraphics[scale=0.30]{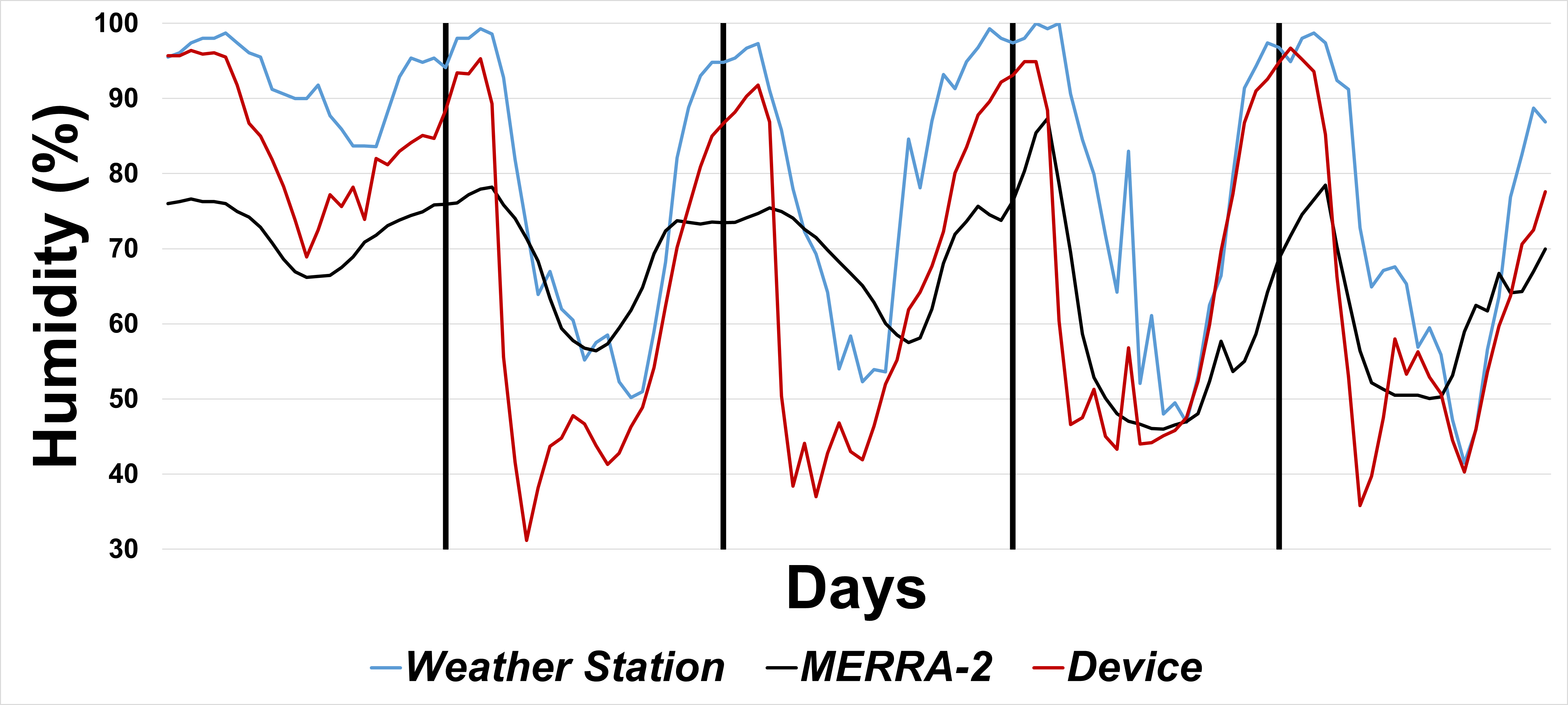}
    \caption{Relative humidity recorded during the Ulm
             deployment, compared against DWD and MERRA-2
             reference data.}
    \label{fig:Humidity}
\end{figure}

\begin{figure}[H]
    \centering
    \includegraphics[scale=0.30]{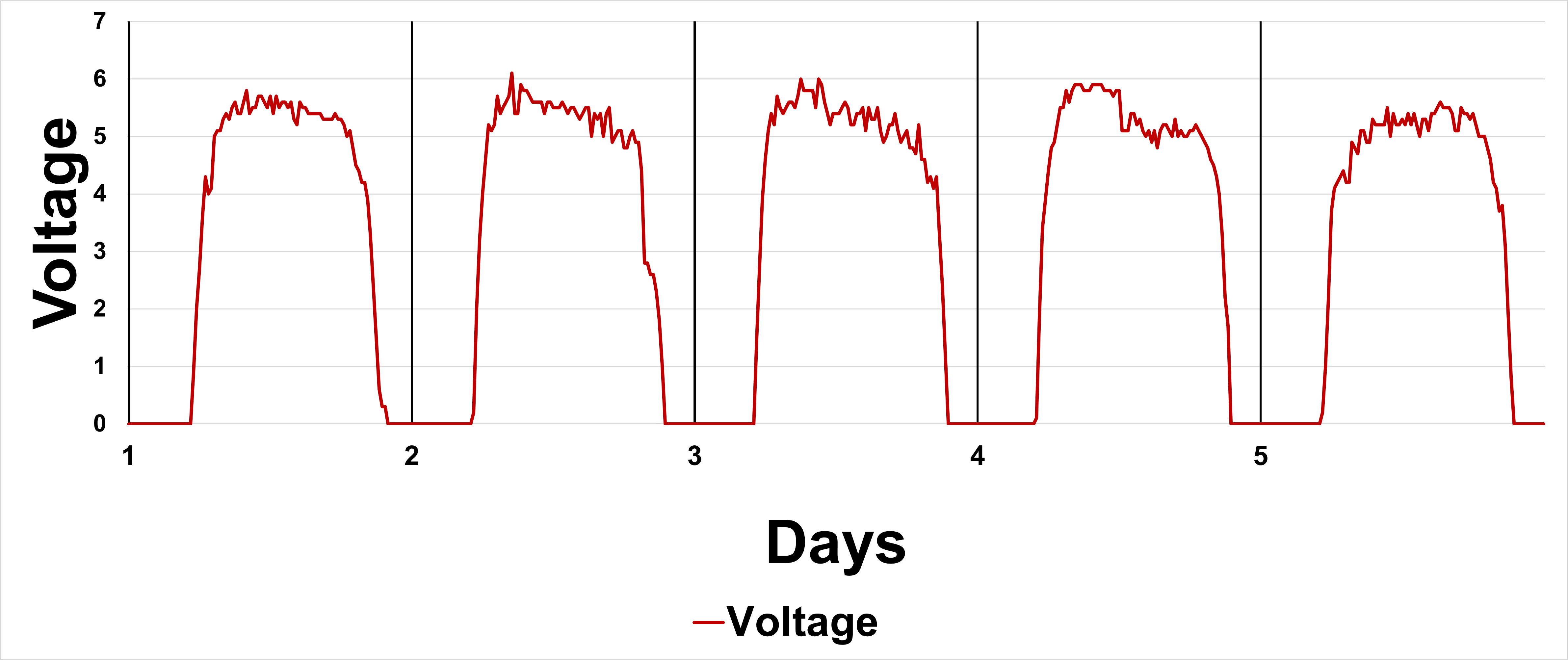}
    \caption{Solar panel voltage recorded during the Ulm
             training period.}
    \label{fig:Voltage}
\end{figure}

Figures~\ref{fig:Temperature}--\ref{fig:Voltage} present the
temporal evolution of the monitored variables throughout the
deployment. Temperature (Figure~\ref{fig:Temperature}) shows
a diurnal pattern consistent with the summer season in Central
Europe, with recorded values ranging from 10.2\,°C to
35.3\,°C. Differences between the device and the reference
datasets are visible over the deployment period, reflecting
the localized thermal conditions of the rooftop installation
that are not resolved at the spatial scale of gridded
reanalysis data \cite{Merra2}. Relative humidity
(Figure~\ref{fig:Humidity}) follows the general diurnal
variation observed in the reference data, with values
ranging between 28.5\% and 99.5\% over the deployment
period. Solar panel voltage (Figure~\ref{fig:Voltage}) shows
values ranging between 0\,V at night and a maximum of 6.4\,V
during peak irradiance hours, with day-to-day variability
reflecting changing atmospheric conditions over the
deployment period.

Solar panel voltage during the training period is shown in
Figure~\ref{fig:Voltage}. Values ranged between 0\,V at
night and a maximum of 6.4\,V during peak irradiance hours,
with day-to-day variability reflecting changing atmospheric
conditions over the deployment period.

Two aspects of this deployment are noted as constraints on
forecasting accuracy. First, the LDR operated in a near-binary
mode at this site, limiting the information available to the
model about irradiance amplitude variations during the day.
Second, the short training period provided a limited sample
of the meteorological variability of the site. Both
observations were taken into account in the design of the
second deployment, which adopted an extended training period
of 84 days.

\subsection{Case Study II: Zapopan, Mexico}

\subsubsection{Experimental Setup and Location}

\begin{figure}[ht]
    \centering
    \includegraphics[width=0.6\linewidth]{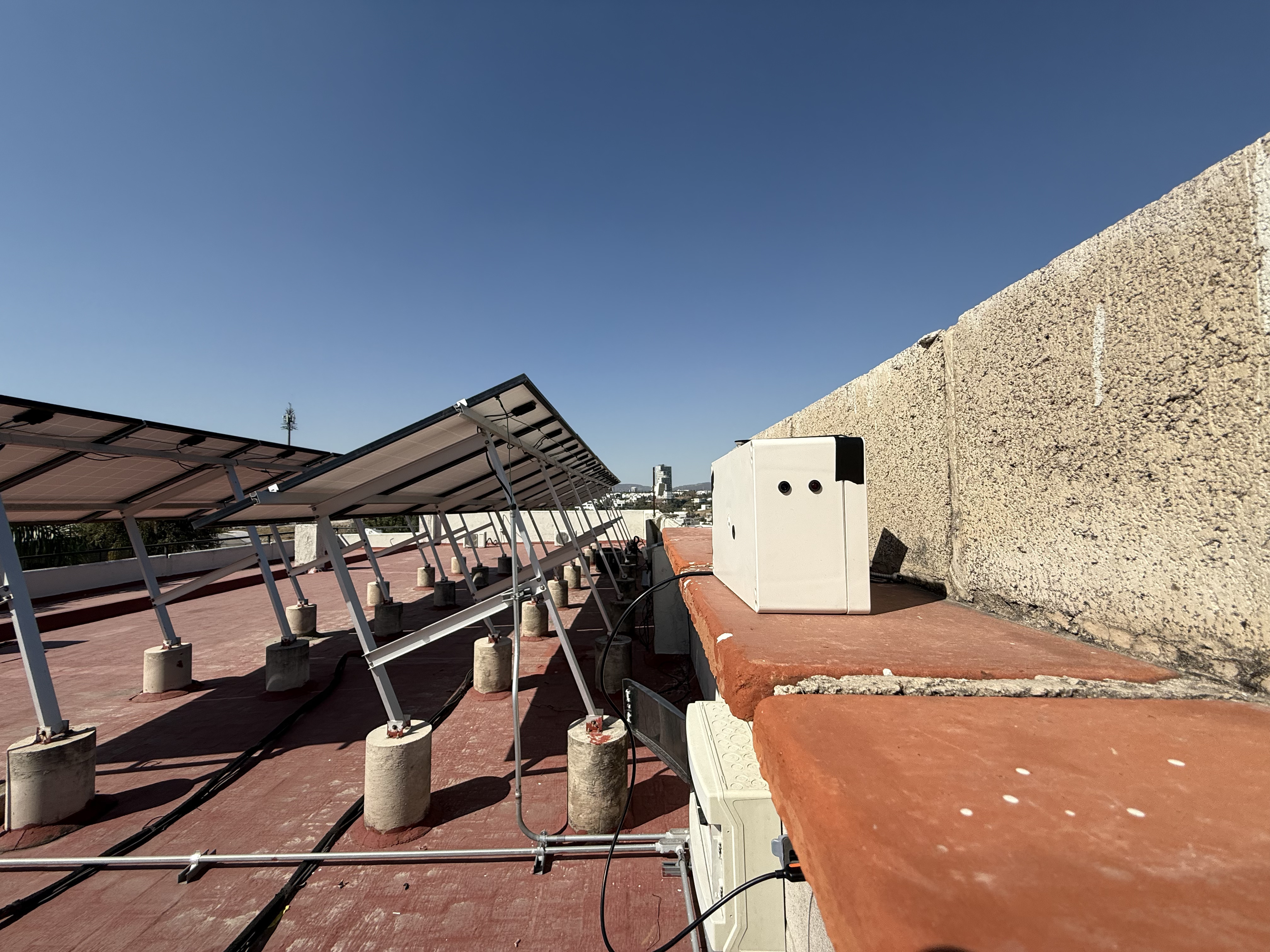}
    \caption{The assembled device deployed on the rooftop of
             Building~J at the Universidad Aut\'onoma de Guadalajara,
             Zapopan, Mexico, during the 115-day field campaign.}
    \label{fig:zapopan_deploy}
\end{figure}

The second deployment was conducted on the rooftop of
Building~J (\textit{Edificio~J}) at the Universidad Aut\'onoma
de Guadalajara, Zapopan, Jalisco, Mexico, from February 6
to May 31, 2025. The rooftop location provided direct solar
exposure throughout the deployment period. The site is
characterized by a subtropical climate with higher solar
irradiance and a broader thermal range compared to the German
site, with temperatures recorded between 8.2\,°C and
53.0\,°C over the full deployment period.
Figure~\ref{fig:zapopan_deploy} shows the device installed at
the Building~J rooftop site.

The 115-day deployment was organized in two phases. The
training phase (February 6 -- April 30, 2025; 84 days)
covered the local dry season and the beginning of the
transition toward higher humidity conditions, providing a
diverse set of environmental conditions for model training.
The validation phase (May 1--31, 2025; 31 days) operated the
embedded feedforward model in fully autonomous predictive mode, generating
24-hour solar voltage forecasts without external retraining
or connectivity. Reference data were obtained from the NASA
MERRA-2 reanalysis \cite{Merra2}; local meteorological
station data were not available for this site during the
experimental period.

\subsubsection{Recorded Environmental Data: Training Phase}

\begin{figure}[H]
    \centering
    \includegraphics[width=0.95\linewidth]{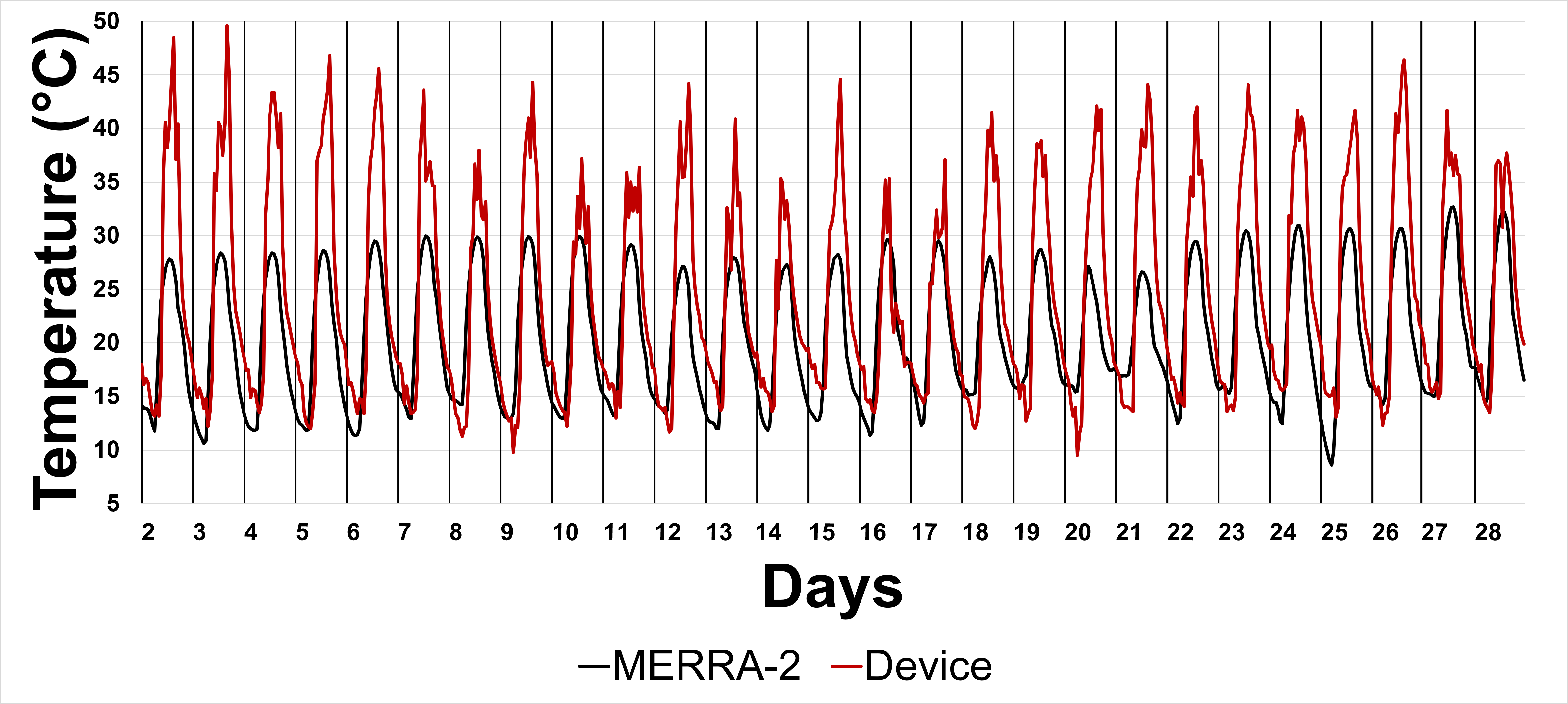} \\
    \textit{(a)~February 2025} \\
    \vspace{0.1cm}
    \includegraphics[width=0.95\linewidth]{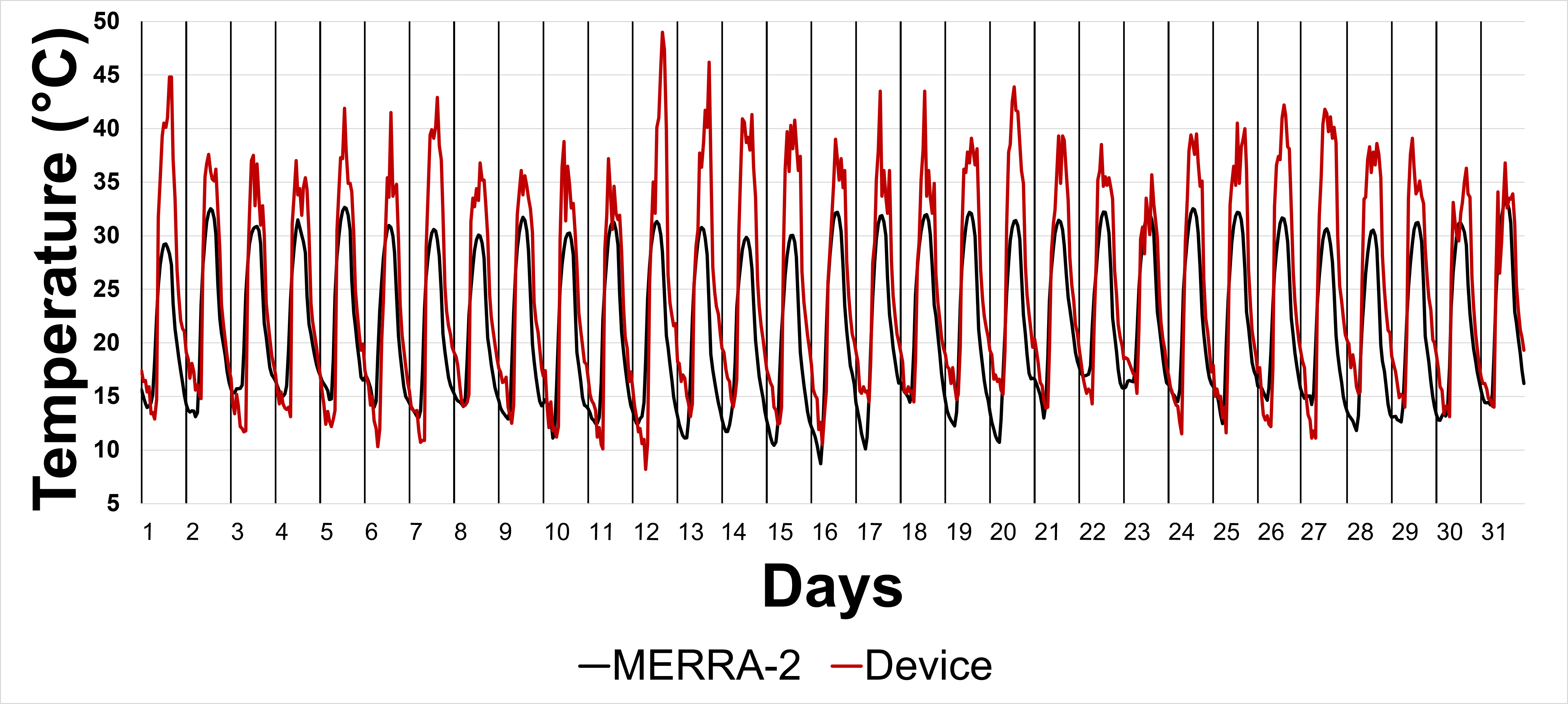} \\
     \textit{(b)~March 2025} \\
\vspace{0.1cm}
    \includegraphics[width=0.95\linewidth]{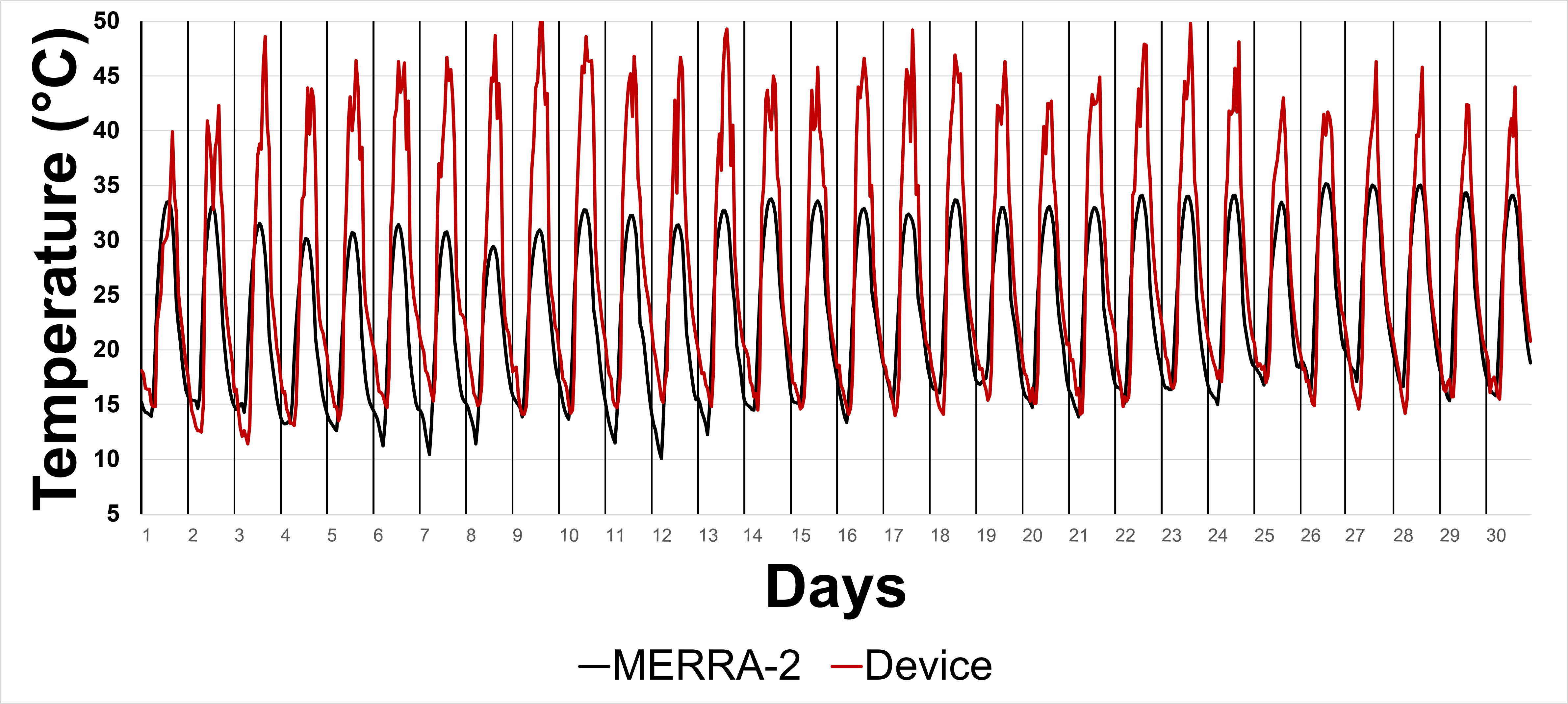} \\
     \textit{(c)~April 2025}
    \caption{Temperature recorded during the Zapopan training
             phase, compared against MERRA-2 reanalysis.}
    \label{fig:TempMex}
\end{figure}

Temperature during the training phase is presented in
Figure~\ref{fig:TempMex} for each of the three training
months. The recorded values reflect the thermal
characteristics of the subtropical site, with temperatures
ranging from 8.2\,°C to 53.0\,°C over the full period.
The MERRA-2 reanalysis values differ from the device
measurements, consistent with the spatial resolution of the
reanalysis grid and the localized thermal conditions of a
rooftop installation \cite{Merra2}.

\begin{figure}[H]
        \centering
    \includegraphics[width=0.95\linewidth]{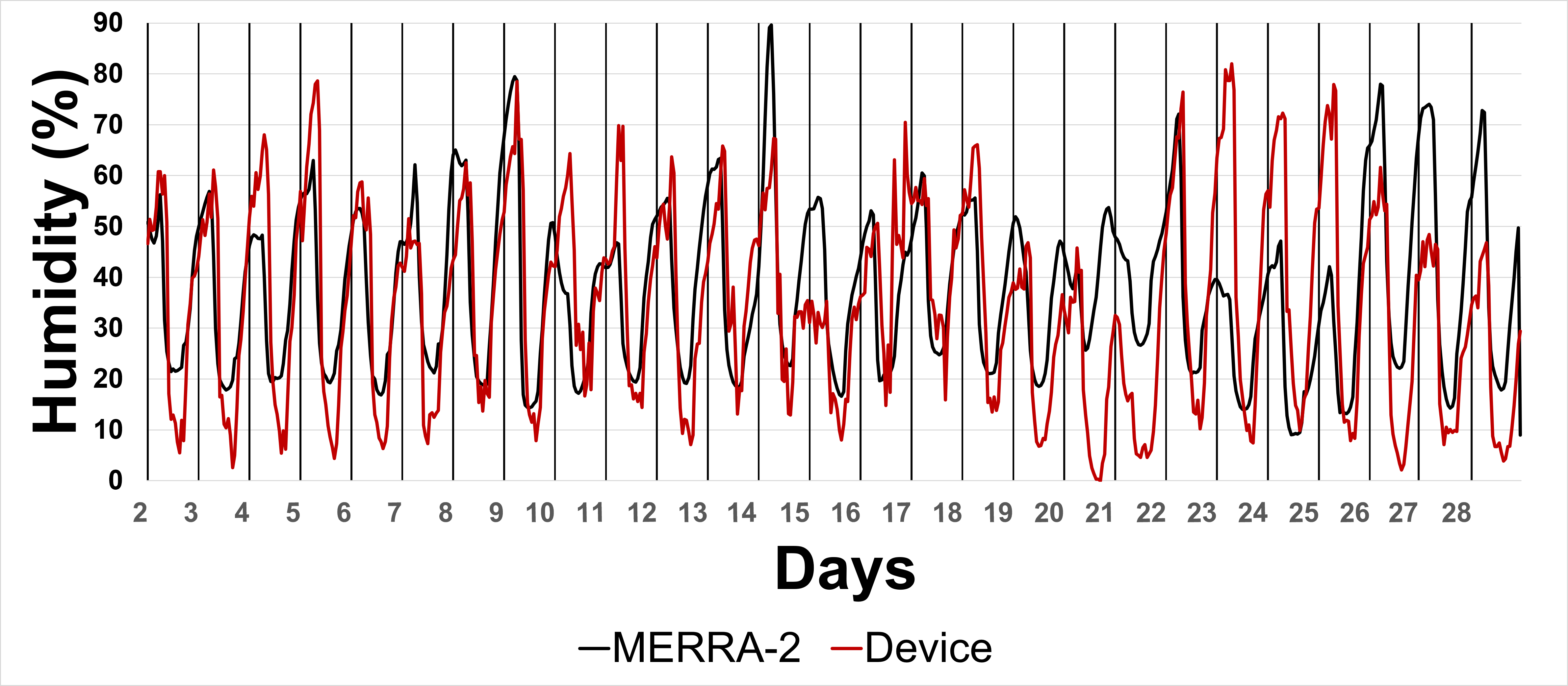} \\
    \textit{(a)~February 2025} \\
    \vspace{0.1cm}
    \includegraphics[width=0.95\linewidth]{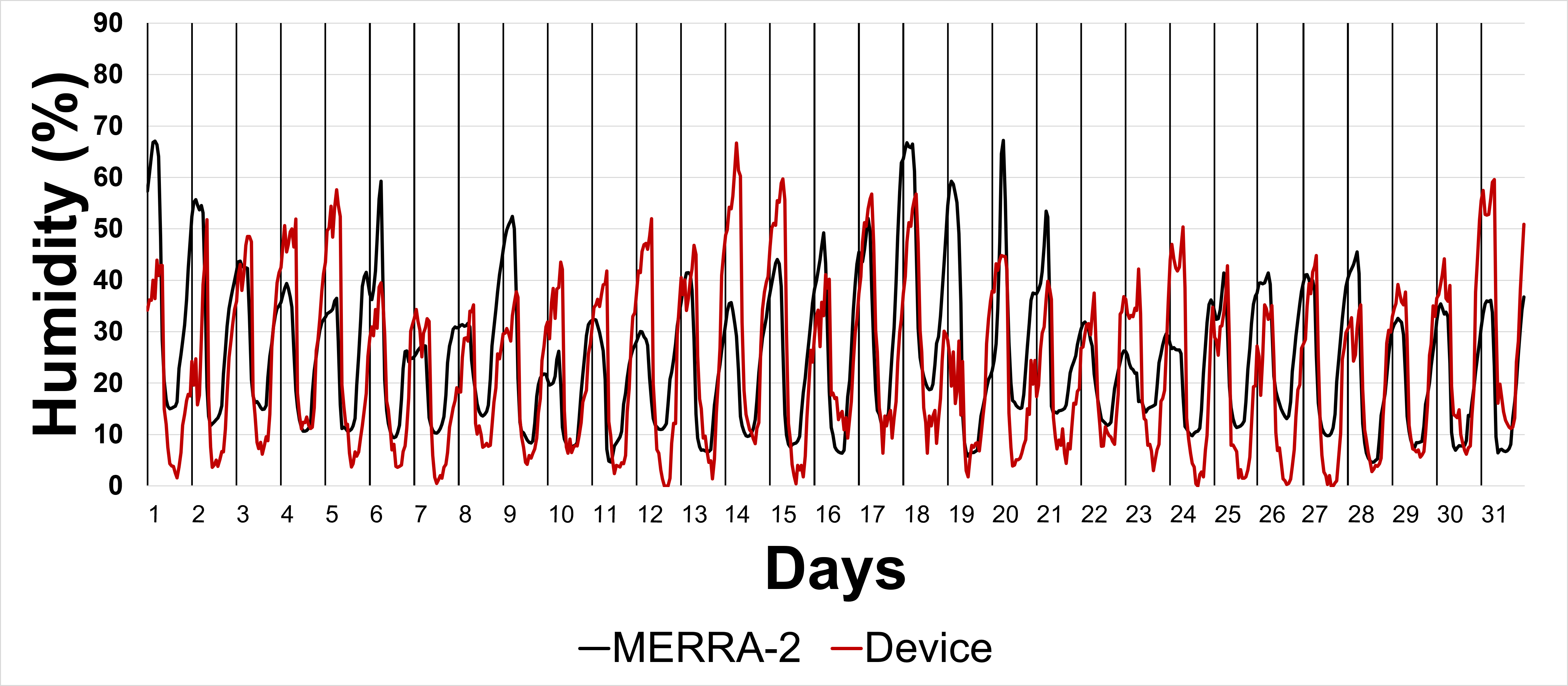} \\
    \textit{(b)~March 2025.} \\
    \vspace{0.1cm}
    \includegraphics[width=0.95\linewidth]{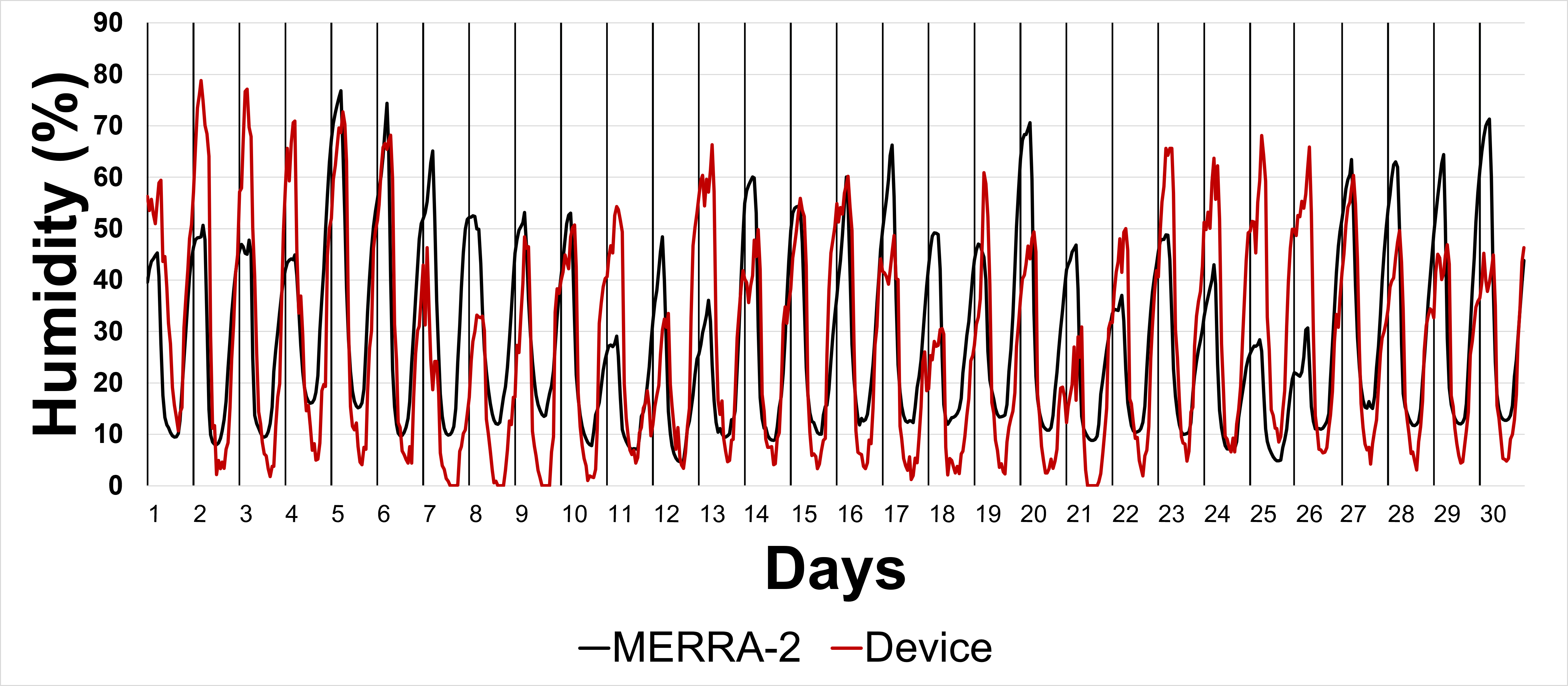} \\
    \textit{(c)~April 2025.}
    \caption{Relative humidity recorded during the Zapopan
             training phase, compared against MERRA-2 reanalysis.}
    \label{fig:HumMex}
\end{figure}

Relative humidity during the training phase is shown in
Figure~\ref{fig:HumMex}. The data reflect the humidity
conditions across the three training months, including an
increase in atmospheric moisture observed toward the end
of April as the region begins its seasonal transition.

\begin{figure}[H]
    \centering
    \includegraphics[width=0.95\linewidth]{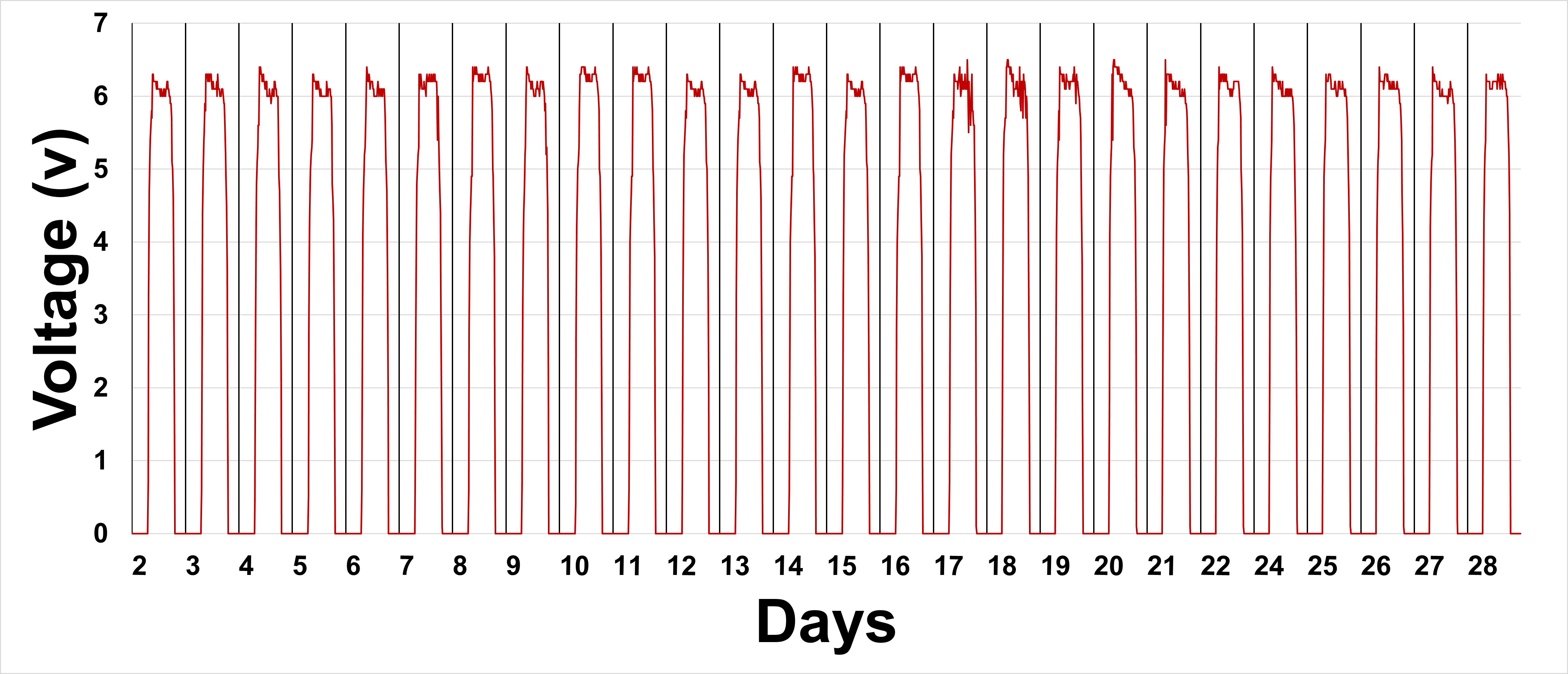} \\
    \textit{(a)~February 2025} \\
    \vspace{0.1cm}
    \includegraphics[width=0.95\linewidth]{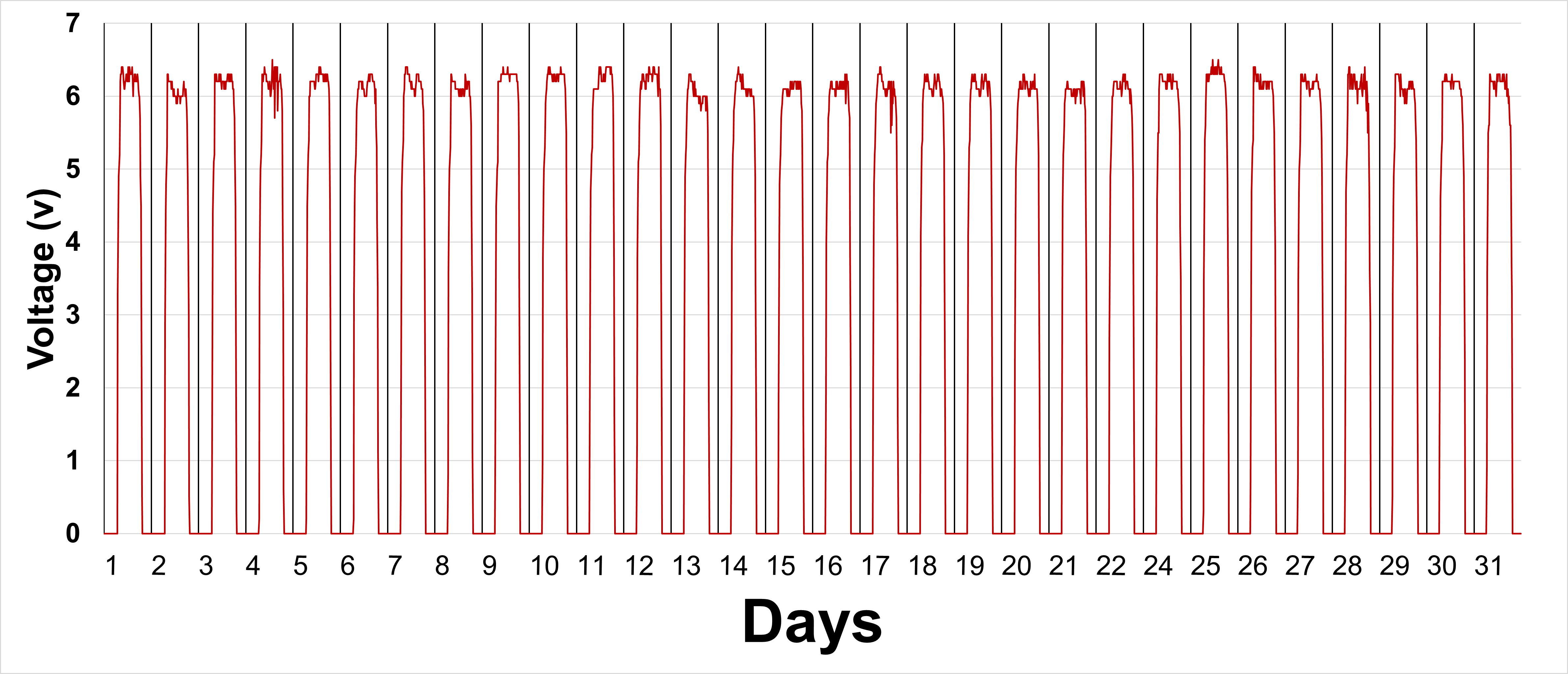} \\
    \textit{(b)~March 2025} \\
     \vspace{0.1cm}
    \includegraphics[width=0.95\linewidth]{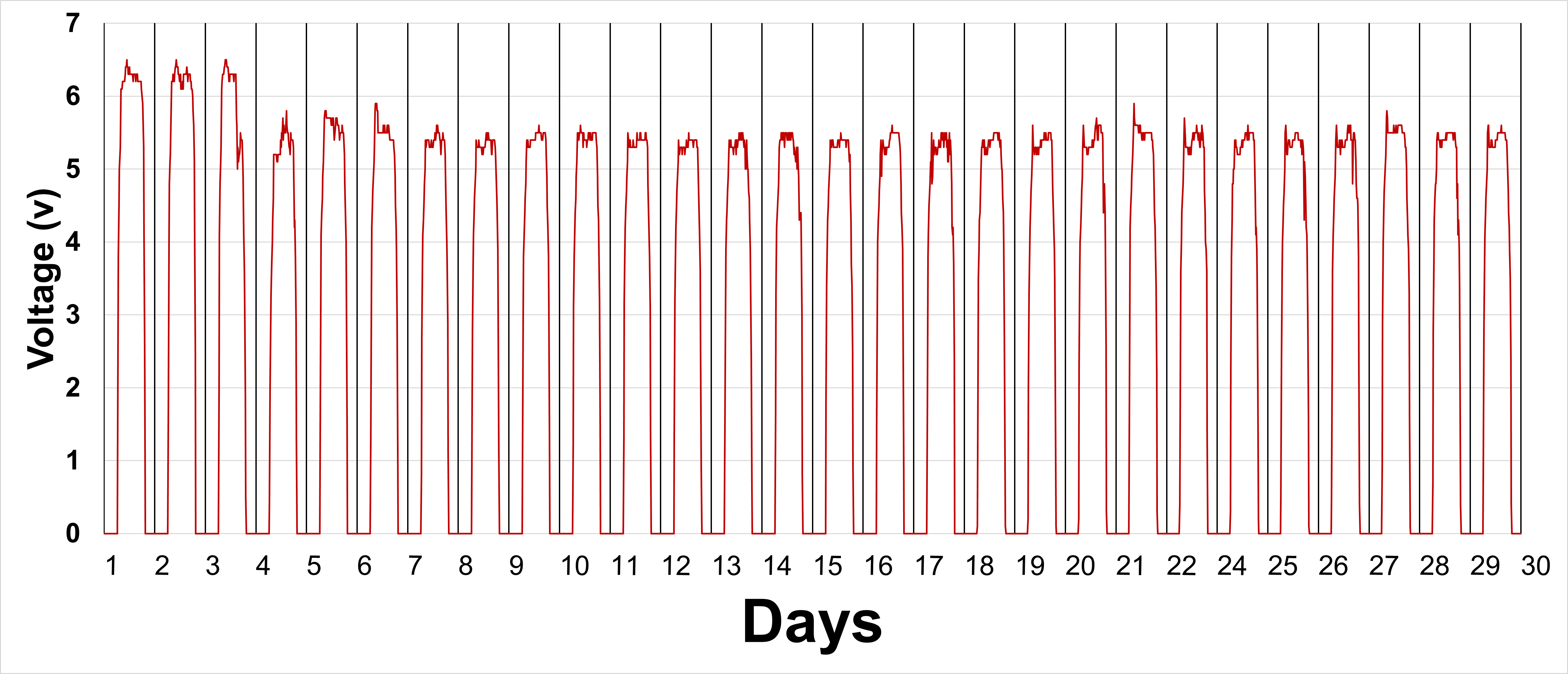} \\
    \textit{(c)~April 2025} \\
    \caption{Solar panel voltage recorded during the Zapopan
             training phase. Peak voltages in February and March
             reached 6.4\,V.}
    \label{fig:VoltMex}
\end{figure}

Solar panel voltage during the training phase is presented
in Figure~\ref{fig:VoltMex}. Peak voltages in February and
March consistently reached the operational ceiling of
6.4\,V, while April recorded lower peak values in the range
of 5.5--5.6\,V, coinciding with the highest recorded
temperatures and increased atmospheric humidity of the
deployment. This variation was captured by the device
sensors and is reflected in the training data used by the
embedded model.

\subsubsection{Forecasting Results}

\begin{figure}[H]
    \centering
    \includegraphics[width=0.9\linewidth]{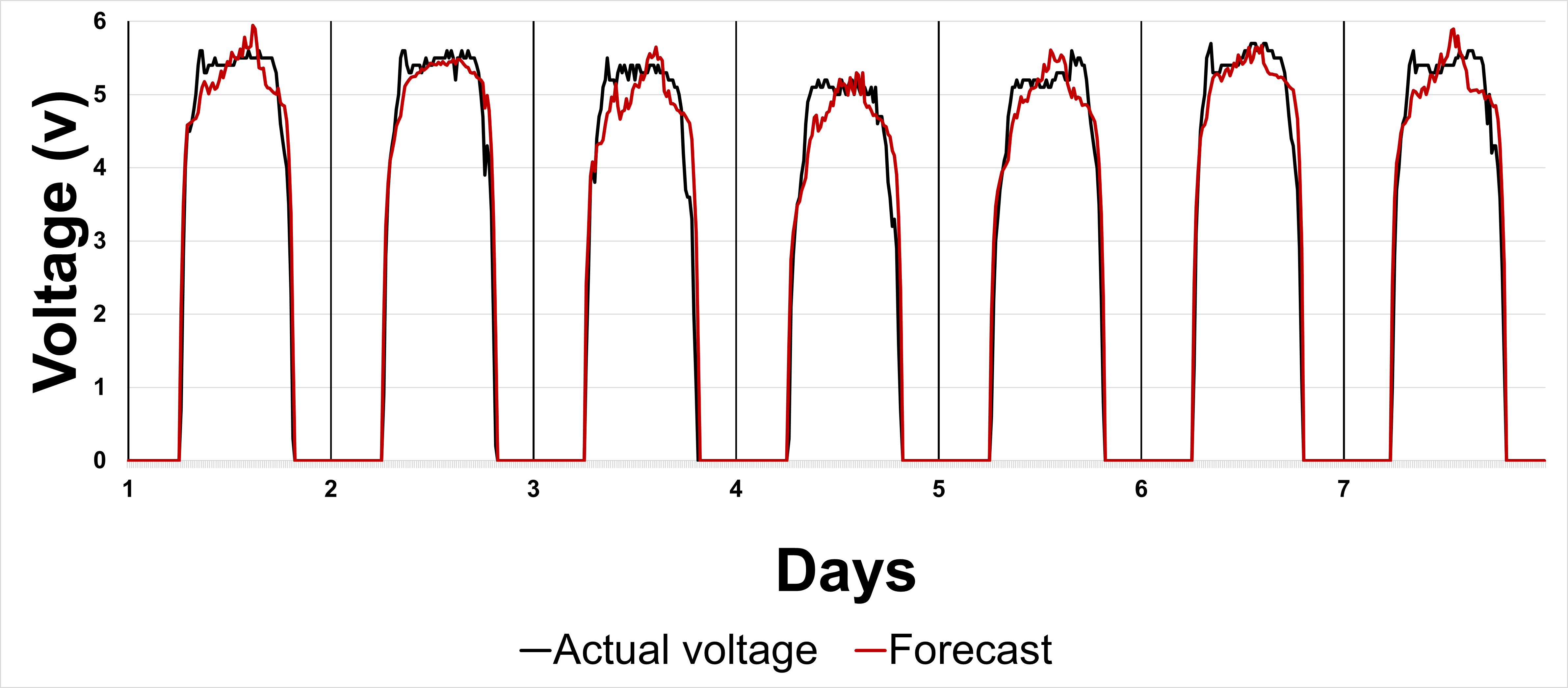}\\
    \textit{ (a)~Week~1 (days~1--7)}\\
    \includegraphics[width=0.9\linewidth]{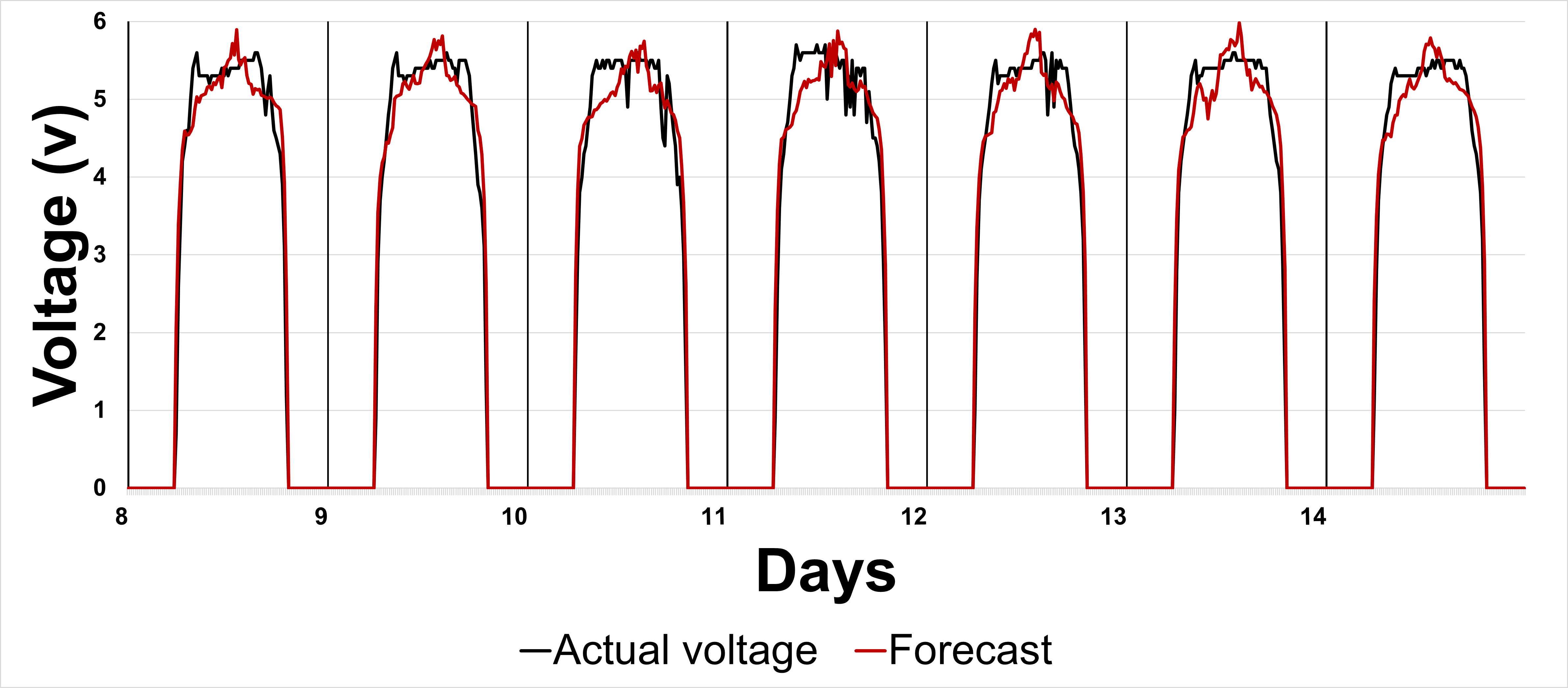}\\
    \textit{(b)~Week~2 (days~8--14)}\\
    \includegraphics[width=0.9\linewidth]{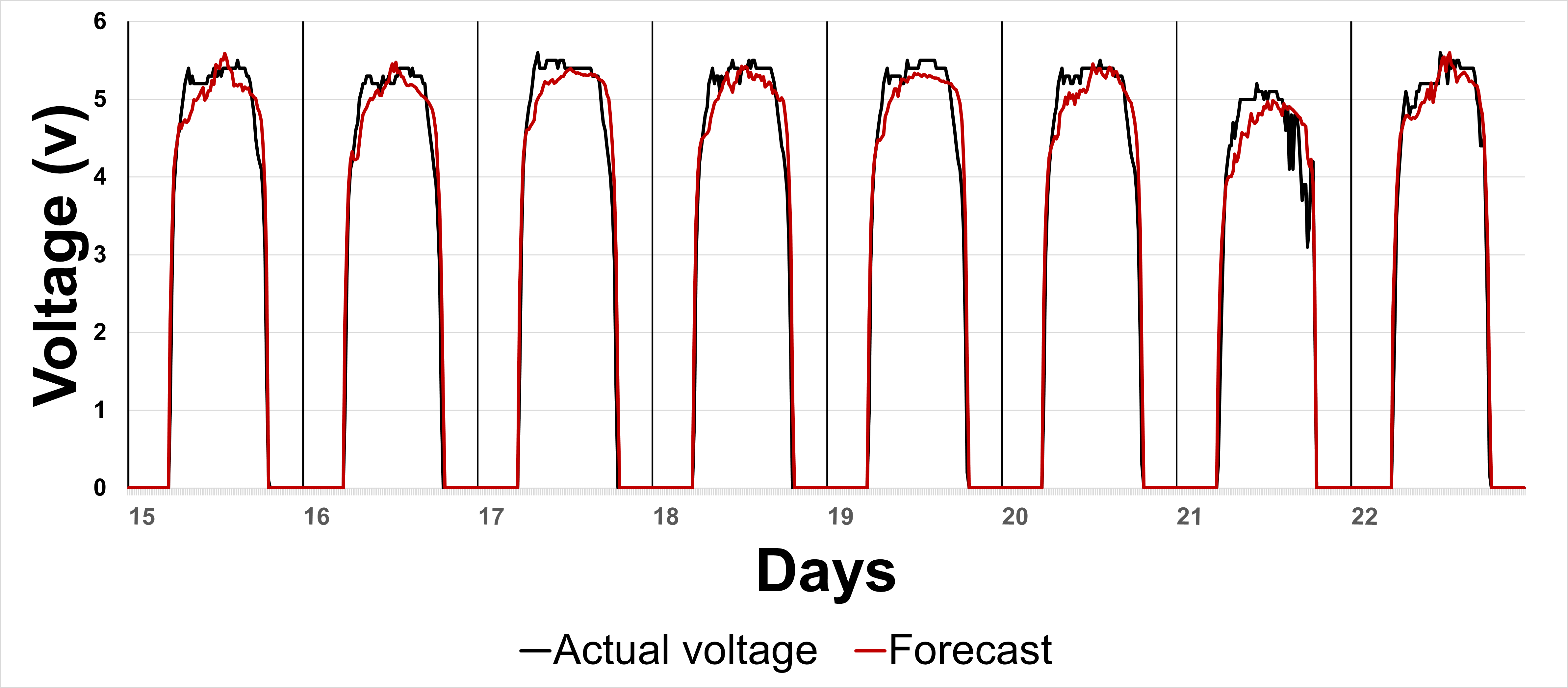}\\
    \textit{(c)~Week~3 (days~15--21)}\\
    \includegraphics[width=0.9\linewidth]{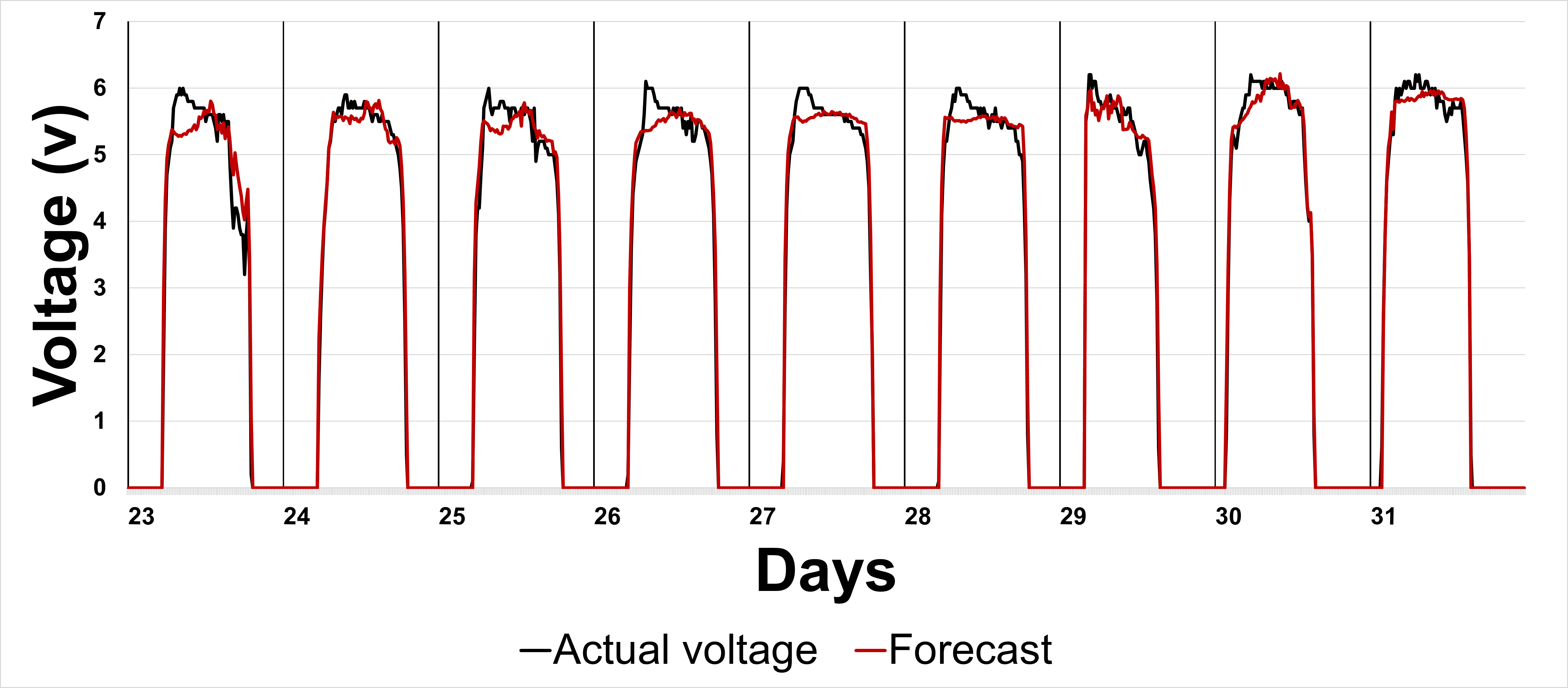}\\
    \textit{(d)~Week~4 (days~22--31).}
    \caption{Predicted vs.\ measured solar panel voltage,
             Zapopan validation phase (May 2025).}
    \label{fig:PredMex}
\end{figure}

The feedforward network reaches convergence during external training without signs of overfitting. The embedded model was then deployed on the ESP32 for the 31-day validation phase of
May 2025. Figure~\ref{fig:PredMex} presents the comparison between predicted and measured solar panel voltage across the four weeks of the validation period.

During Week~1 (Figure~\ref{fig:PredMex}a), the model
produced the largest deviations of the validation period,
with errors in peak voltage magnitude reaching 0.3--0.4\,V
on the most challenging days. The model reproduced
the diurnal shape, while the zero-voltage outputs during
nighttime intervals are imposed by a firmware rule
(Section~\ref{sec:embedded}) rather than predicted by the network.

Over the following weeks (Figures~\ref{fig:PredMex}b--d),
the mean daytime error decreased from 0.347\,V in Week~1 to
0.227\,V in Week~4, a reduction of approximately 35\%.
Part of this improvement reflects the on-device weight update
(Eq.~\ref{eq:update}) incorporating each preceding day's
measurements through the sliding window
$\mathcal{W}_t$ (Eq.~\ref{eq:window_def}); however, as the
frozen-weight ablation in Section~\ref{sec:adaptation} shows,
a comparable trend is present even without weight updates,
indicating that the declining error is driven in large part
by the May meteorological regime progressively resembling the
training distribution rather than by adaptation alone.
The final three days of May (29--31) are excluded from all
quantitative metrics: an on-device real-time-clock reset
shifted their timestamps, placing the daily voltage peak near
05:00 instead of the physical solar noon, which would
otherwise corrupt the time-aligned error computation.

The aggregate performance metrics are reported in
Table~\ref{tab:metrics}. All metrics are computed over a clean
evaluation window comprising the 28 unaffected validation days
(1--28~May) restricted to daytime hours (06:00--19:00), which
excludes both the RTC-corrupted final days and the nighttime
intervals whose zero output is fixed in firmware. Over this
window the model attains a mean absolute error (MAE) of
0.2975\,V, corresponding to 4.65\% of the full operational
voltage range (0--6.4\,V), with a small positive bias of
0.0225\,V (0.35\% of range) and a coefficient of determination
$R^2 = 0.9165$. Table~\ref{tab:baselines} places these figures
against two reference forecasts: the model improves substantially
on a climatology baseline (skill score $+0.64$) but does not
surpass a 24-hour persistence baseline (skill score $-0.39$),
a strong benchmark on clear-sky dry-season series with high
day-to-day autocorrelation. The regime-dependence of this
comparison is examined in Section~\ref{sec:trainingEffect}.

\begin{table}[ht!]
\caption{Forecasting performance of the embedded feedforward model,
         Zapopan validation phase. Metrics computed over the clean
         evaluation window: 28 validation days (1--28~May~2025),
         daytime hours only (06:00--19:00), at 15-minute resolution.
         Raw measurements and model predictions are available in the
         public repository \cite{Erick2024}.}
\label{tab:metrics}
\centering
\begin{tabular}{lll}
\toprule
\textbf{Metric} & \textbf{Value} & \textbf{Interpretation} \\
\midrule
Mean Absolute Error (MAE)
    & 0.2975\,V
    & 4.65\% of operational range \\
Mean Error (bias)
    & $+$0.0225\,V
    & $+$0.35\% of operational range \\
Coefficient of determination ($R^2$)
    & 0.9165
    & 91.65\% of variance explained \\
Operational voltage range
    & 0--6.4\,V
    & After voltage divider \\
\bottomrule
\end{tabular}
\end{table}

\begin{table}[ht!]
\caption{Comparison of the embedded model against two reference
         forecasts over the clean evaluation window (28 daytime
         days). The skill score is $1 - \mathrm{MAE}_{\mathrm{model}}
         / \mathrm{MAE}_{\mathrm{reference}}$; positive values
         indicate the model outperforms the reference.}
\label{tab:baselines}
\centering
\begin{tabular}{lll}
\toprule
\textbf{Forecast} & \textbf{MAE (V)} & \textbf{Skill vs.\ model} \\
\midrule
Embedded model        & 0.2975 & --- \\
Persistence (24\,h)   & 0.2142 & $-0.39$ \\
Climatology           & 0.8223 & $+0.64$ \\
\bottomrule
\end{tabular}
\end{table}
 
\section{Discussion}
\label{sec:discussion}

\subsection{Comparison with related work}
\label{sec:related}

Several prior systems have deployed machine learning on low-cost
microcontrollers for environmental or photovoltaic monitoring.
Samara and Natsheh \cite{samara2019intelligent} implemented an
artificial neural network for PV panel monitoring on a
microcontroller, and Sridhar et al.\ \cite{SRIDHAR2023100609}
proposed a modular ESP32 platform for on-device air-quality
inference; both perform inference on-device but rely on models
trained once and fixed thereafter. Melo et al.\ \cite{Melo221}
implemented real-time PV monitoring with local data storage but
without an embedded forecasting model. Ahmed et al.\
\cite{Ahmed2022} surveyed edge-based solar forecasting and
identified the deployment of adaptive models on
resource-constrained microcontrollers as an open problem.
Against this background, the distinguishing feature of the present
work is not forecasting accuracy, which, as shown below, does not
surpass a persistence baseline, but the combination of fully
on-device inference with a post-deployment incremental update
mechanism operating without cloud connectivity, evaluated over a
115-day field campaign with an explicit baseline comparison and
ablation.

\subsection{Training duration and sensor constraints}
\label{sec:trainingEffect}

The contrast between the two field deployments indicates that
training duration is an important factor governing forecasting
accuracy. In Ulm, training covered only a short period, and the
model's capacity to resolve intra-day irradiance amplitude
variations was further constrained by the near-binary signal of
the LDR under the diffuse sky conditions predominant at the site;
no quantitative error metrics are reported for that deployment,
which served as a hardware and firmware validation rather than a
forecasting benchmark. In Zapopan, an 84-day
training period spanning the complete local dry season and the
onset of the humidity transition exposed the model to the full
range of site meteorological variability, including the systematic
afternoon voltage depression observed in late April at high
atmospheric humidity. The result was a mean absolute error of
0.2975\,V (4.65\% of the operational voltage range) across the
clean daytime evaluation window. Because the two deployments differ
in sensor behaviour, season, and evaluation protocol, they cannot
be compared quantitatively; the Zapopan campaign is the only one
from which forecasting accuracy can be assessed, and the effect of
training duration is therefore best regarded as a design
consideration supported by the Ulm experience rather than as a
measured contrast.

\subsection{On-device adaptation and incremental learning}
\label{sec:adaptation}

The mean daytime error across the Zapopan validation month
decreased from 0.347\,V in Week~1 to 0.227\,V in Week~4, a
reduction of approximately 35\%. To determine how much of this
improvement is attributable to the on-device gradient descent
update rather than to a shifting meteorological regime, we
conducted a frozen-weight ablation: the model trained offline
was run over the May inputs both with the daily weight update
active and with the weights held fixed, starting from identical
initial parameters. The experiment was repeated across five random
initializations. The adaptive model achieved a lower mean absolute
error than its frozen counterpart in all five runs, with a mean
improvement of $0.025 \pm 0.006$\,V (paired $t$-test,
$p = 0.001$; Table~\ref{tab:ablation}).

Two conclusions follow. First, the adaptation effect is real and
statistically significant across the five random initializations
tested ($n = 5$): on-device incremental learning does improve
forecasts relative to a static model, without external retraining,
cloud connectivity, or physical access to the device. Second, the
effect is small in absolute terms, about 0.38\% of the operational
range, and the frozen model exhibits a comparable week-to-week
decline in error. The bulk of the observed accuracy gain over the
validation month therefore arises from the May meteorological
conditions progressively resembling the training distribution, not
from adaptation. The value of the update mechanism is thus better
understood as protection against distributional drift, maintaining
a consistent advantage over a static model as conditions evolve, rather
than as a source of large accuracy gains. The choice $\eta = 0.001$
provides this adaptation without destabilizing the weights
established during offline training
\cite{Losing2018incremental, Erick2024}.

\begin{table}[ht!]
\caption{Frozen-weight ablation over the clean daytime evaluation
         window. Mean absolute error of the adaptive model versus a
         frozen-weight model starting from identical initial
         parameters, across five random initializations.}
\label{tab:ablation}
\centering
\begin{tabular}{lll}
\toprule
\textbf{Configuration} & \textbf{MAE (V)} & \textbf{} \\
\midrule
Frozen weights   & $0.514 \pm 0.055$ & baseline \\
Adaptive update  & $0.489 \pm 0.050$ & $-0.025$\,V ($p = 0.001$) \\
\bottomrule
\end{tabular}
\end{table}

\subsection{Limitations and generalizability}

Three limitations bound the scope of the conclusions presented here.
First, each deployment site was observed across a single seasonal
window: Ulm over late spring (May--June 2024) and Zapopan across
the dry-season-to-wet-season transition (February--May 2025).
Within-site generalization across all seasons, and in particular
the model's performance under conditions substantially different
from those in the training set, remains to be demonstrated;
the $R^2 = 0.9165$ reported for Zapopan is specific to May 2025
conditions and should not be extrapolated to, for example, the
full rainy season. Second, the LDR GL5528 provides only an
uncalibrated 0--100 relative light index rather than
spectrally calibrated irradiance in W/m$^2$; its near-binary
behavior under diffuse radiation limits the amplitude information
available to the model on overcast days, which is the principal
sensor-level constraint on forecasting fidelity under broken-cloud
conditions. Third, the fundamental accuracy ceiling of any
single-site, purely local monitoring system is set by meteorological
phenomena at synoptic scale, such as pressure-front passages in Ulm and
Pacific moisture intrusions in Zapopan, that cannot be
characterized from ground-level measurements alone. Integrating
regional numerical weather prediction (NWP) fields as additional
input features represents the most productive direction for
addressing this inherent limitation \cite{wan2015photovoltaic,
Merra2}.

\section{Conclusion}
\label{sec:conclusion}

In this article we present a low-cost open-source sensor box for recording climate data. Exemplarily we fitted the box with temperature, humidity, and solar power sensors. However, there is the option to install, say, also a wind sensor. We explain both hardware and software setup and present a forecasting method that is fit to run on the box without internet connection or the necessity for remote (online) calibration. This work demonstrates that a sub-\$55 IoT device
integrating an ESP32 microcontroller with an embedded
feedforward network can generate autonomous 24-hour solar
voltage forecasts on-device, attaining a mean absolute error
of 0.2975\,V (4.65\% of the operational range, $R^2 = 0.9165$)
over a clean 28-day evaluation window in Zapopan. The model
improves substantially on a climatology baseline but does not
surpass a 24-hour persistence baseline, a strong benchmark on
the clear-sky dry-season conditions observed during validation.
A frozen-weight ablation confirms that the on-device incremental
update yields a small but statistically robust accuracy gain
($p = 0.001$), establishing that autonomous incremental learning
is feasible on this class of hardware even where its practical
magnitude is modest. The
hybrid architecture, which decouples computationally intensive
external training from on-device inference via optimized
static weight matrices, resolves the tension
between algorithmic sophistication and the resource
constraints of low-cost embedded hardware, enabling
deployment in contexts where conventional forecasting
infrastructure remains economically inaccessible. The
on-device sliding-window update mechanism allows the
model to adapt to local conditions after deployment without
external retraining or cloud connectivity \cite{Merra2, Erick2024}.

Future work should prioritize three directions. First,
replacing the LDR photoresistor with a quantitative
irradiance sensor would provide continuous amplitude
information during partial cloud conditions, improving
model input quality beyond the near-binary signal observed
at the German site. Second, extending training periods to
cover complete seasonal cycles would strengthen the
model's exposure to the full range of meteorological
variability at each site. Third, transitioning from
Bluetooth to low-power wide-area network protocols such
as LoRaWAN would enable multi-node sensor networks capable
of generating spatially distributed forecasts, making the
system suitable for microgrid operators and regional energy
planning applications \cite{Erick2024}.

\vspace{6pt} 




\textbf{authorcontributions:}{Conceptualization, E.P., A.D. and S.S.; methodology, E.P. and A.D.; software, E.P. and A.D.; validation, S.S. and A.D.; formal analysis, E.P. and A.D.; investigation, E.P.; resources, S.S.; data curation, E.P.; writing---original draft preparation, E.P., A.D. and S.S.; writing---review and editing, A.D. and S.S.; visualization, E.P.; supervision, S.S.; project administration, S.S.; funding acquisition, S.S. All authors have read and agreed to the published version of the manuscript.}

\textbf{funding:}{This research received no external funding.}

\textbf{data availability:}{The raw sensor measurements and model predictions from
both field deployments, together with the firmware source code and the
companion mobile application, are publicly available in the Mendeley Data
repository \cite{Erick2024}.}

\textbf{conflicts of interest:}{The authors declare no conflicts of interest.} 





\appendixpage

\appendix

\section{Sensor Costs}

\begin{table}[ht]
\tiny
    \centering  
    \caption{Approximate Cost Breakdown of the Hardware Components}  
    \label{tab:costs}  
    \renewcommand{\arraystretch}{1.1}
    \setlength{\tabcolsep}{8pt}
    \begin{tabular}{|l|l|r|}  
        \hline  
        \textbf{Component}            & \textbf{Description}                      & \textbf{Approx. Cost (USD)} \\ \hline  
        ESP32-WROOM32                 & Microcontroller with Wi-Fi/Bluetooth      & 15.85     \\ \hline  
        DHT22                         & Temperature and Humidity Sensor           & 4.27 \\ \hline  
        Photoresistor (LDR)           & Light Intensity Sensor                    & 0.93                        \\ \hline  
        Solar Panel (9V, 1.2W)        & Energy Generation Measurement             & 9.23     \\ \hline  
        RTC DS3231                    & Real-Time Clock for Time Synchronization  & 3.42      \\ \hline  
        MicroSD Card Module           & Data Logging Storage Module               & 2.51     \\ \hline  
        Power Bank (10{,}000\,mAh)    & Power Supply for Device                   & 11.39   \\ \hline  
        Protective Enclosure (IP68)   & Waterproof and Dustproof Enclosure        & 10.25    \\ \hline  
       Miscellaneous Components      & Resistors, Capacitors, Wires, Connectors   & 7     \\ \hline  
        \textbf{Total Approximate Cost} &                                          & \textbf{64.83}              \\ \hline  
    \end{tabular}\\[0.25cm]
    Approximate retail unit prices on Amazon Germany,
    accessed 28 July 2026, converted from EUR at a rate of 1.14~USD/EUR.
    These figures represent an upper reference; local sourcing in Mexico
    yields a lower build cost. The power bank and miscellaneous components (resistors,
    capacitors, wires) listed here are operational accessories required for
    field deployment and are not part of the eight core sensing and
    processing components enumerated in Table~\ref{tab:bom}; they are
    included in this breakdown only to give a complete estimate of the
    total cost of a field-ready unit.
\end{table}







\bibliographystyle{elsarticle-num} 
\bibliography{bibfile}

%


\end{document}